\documentclass[aps,prx,preprintnumbers,floats,twocolumn,superscriptaddress,nofootinbib]{revtex4}
\usepackage{graphicx}% Include figure files
\usepackage{dcolumn}% Align table columns on decimal point
\usepackage{bm}% bold math
\usepackage[english]{babel}
\usepackage[latin1]{inputenc}
\usepackage{graphicx}
\usepackage{epstopdf}
\usepackage{amsfonts}
\usepackage{comment}
\usepackage{url}
\usepackage{hyperref}
\usepackage{color}
\usepackage{epsfig}
\usepackage{mathrsfs}
\usepackage{amsmath,amssymb}
\usepackage{subfigure}

\usepackage{framed}
\usepackage{lipsum}
\usepackage[svgnames]{xcolor}
\definecolor{shadecolor}{named}{LightGrey}
\newcommand{\lto}[1]{\longrightarrow#1}

\renewcommand{\(}{\left(}
\renewcommand{\)}{\right)}
\renewcommand{\[}{\left[}
\renewcommand{\]}{\right]}

\begin{document}

\graphicspath{{figure/}}
\selectlanguage{english}

%%%%%%%%%%%%%%%%%%

\title{Mathematical formulation of multi-layer networks}

\author{Manlio De Domenico}
 \affiliation{Departament d'Enginyeria Inform\`atica i Matem\`atiques, Universitat Rovira i Virgili, 43007 Tarragona, Spain}

\author{Albert Sol\'e-Ribalta}
 \affiliation{Departament d'Enginyeria Inform\`atica i Matem\`atiques, Universitat Rovira i Virgili, 43007 Tarragona, Spain}

\author{Emanuele Cozzo}
 \affiliation{Institute for Biocomputation and Physics of Complex Systems (BIFI), University of Zaragoza, Zaragoza 50018, Spain}

\author{Mikko Kivel\"a}
 \affiliation{Oxford Centre for Industrial and Applied Mathematics, Mathematical Institute, University of Oxford, Oxford OX1 3LB, UK}

\author{Yamir Moreno}
 \affiliation{Institute for Biocomputation and Physics of Complex Systems (BIFI), University of Zaragoza, Zaragoza 50018, Spain}
 \affiliation{Department of Theoretical Physics, University of Zaragoza, Zaragoza 50009, Spain}
\affiliation{Complex Networks and Systems Lagrange Lab, Institute for Scientific Interchange, Turin, Italy}

\author{Mason A. Porter}
 \affiliation{Oxford Centre for Industrial and Applied Mathematics, Mathematical Institute and CABDyN Complexity Centre, University of Oxford, Oxford OX1 3LB, UK}

\author{Sergio G\'omez}
 \affiliation{Departament d'Enginyeria Inform\`atica i Matem\`atiques, Universitat Rovira i Virgili, 43007 Tarragona, Spain}

\author{Alex Arenas}
 \affiliation{Departament d'Enginyeria Inform\`atica i Matem\`atiques, Universitat Rovira i Virgili, 43007 Tarragona, Spain}

\date{\today}

\begin{abstract}

A network representation is useful for describing the structure of a large variety of complex systems. However, most real and engineered systems have multiple subsystems and layers of connectivity, and the data produced by such systems is very rich.  Achieving a deep understanding of such systems necessitates generalizing ``traditional" network theory, and the newfound deluge of data now makes it possible to test increasingly general frameworks for the study of networks.  In particular, although adjacency matrices are useful to describe traditional single-layer networks, such a representation is insufficient for the analysis and description of multiplex and time-dependent networks.  One must therefore develop a more general mathematical framework to cope with the challenges posed by multi-layer complex systems. In this paper, we introduce a tensorial framework to study multi-layer networks, and we discuss the generalization of several important network descriptors and dynamical processes---including degree centrality, clustering coefficients, eigenvector centrality, modularity, Von Neumann entropy, and diffusion---for this framework.  We examine the impact of different choices in constructing these generalizations, and we illustrate how to obtain known results for the special cases of single-layer and multiplex networks.  Our tensorial approach will be helpful for tackling pressing problems in multi-layer complex systems, such as inferring who is influencing whom (and by which media) in multichannel social networks and developing routing techniques for multimodal transportation systems.

\end{abstract}

\maketitle

\flushbottom

%%%%%%%%%%%%

%%%%%%%%%%%%%%%%%%%%%%%%%%%%%%%%%%%%%
%%%%%%%%%%%%%%%%%%%%%%%%%%%%%%%%%%%%%
%%%%%%%%%%%%%%%%%%%%%%%%%%%%%%%%%%%%%

%%%%%%%%

\section{Introduction}

The quantitative study of networks is fundamental for the study of complex systems throughout the biological, social, information, engineering, and physical sciences \cite{bocca2006,newman2010,estrada2011}.  The broad applicability of networks, and their success in providing insights into the structure and function of both natural and designed systems, has thus generated considerable excitement across myriad scientific disciplines.  For example, networks have been used to represent interactions between proteins, friendships between people, hyperlinks between Web pages, and much more.  Importantly, several features arise in a diverse variety of networks.  For example, many networks constructed from empirical data exhibit heavy-tailed degree distributions, the small-world property, and/or modular structures; such structural features can have important implications for information diffusion, robustness against component failure, and many other considerations \cite{bocca2006,newman2010,estrada2011}.

Traditional studies of networks generally assume that nodes are connected to each other by a single type of static edge that encapsulates all connections between them.  This assumption is almost always a gross oversimplification, and it can lead to misleading results and even the sheer inability to address certain problems.  For example, ignoring time-dependence throws away the ordering of pairwise human contacts in transmission of diseases \cite{holme2012}, and ignoring the presence of multiple types of edges (which is known as ``multiplexity" \cite{faust}) makes it hard to take into account the simultaneous presence and relevance of multiple modes of transportation or communication.

Multiplex networks explicitly incorporate multiple channels of connectivity in a system, and they provide a natural description for systems in which entities have a different set of neighbors in each layer (which can represent, e.g., a task, an activity, or a category). A fundamental aspect of describing multiplex networks is defining and quantifying the interconnectivity between different categories of connections. This amounts to switching between layers in a multi-layer system, and the associated inter-layer connections in a network are responsible for the emergence of new phenomena in multiplex networks. Inter-layer connections can generate new structural and dynamical correlations between components of a system, so it is important to develop a framework that takes them into account. Note that multiplex networks are not simply a special case of interdependent networks \cite{gao2011networks}: in multiplex systems, many or even all of the nodes have a counterpart in each layer, so one can associate a vector of states to each node.
For example, a person might currently be logged into Facebook (and hence able to receive information there) but not logged into some other social networking site.  The presence of nodes in multiple layers of a system also entails the possibility of self-interactions.
This feature has no counterpart in interdependent networks, which were conceived as interconnected communities within a single, larger network \cite{buldyrev2010,dickison2012epidemics}.

%%%%%%

The scientific community has been developing tools for temporal networks for several years \cite{holme2012,holme2013}, though much more work remains to be done, and now an increasingly large number of scholars with diverse expertise have turned their attention to studying multiplex networks (and related constructs, such as the aforementioned interdependent networks and so-called ``networks of networks")\cite{TTT,raissa2009,mucha2010community,blasius2010,criado2010,buldyrev2010,gao2011,gao2011,raissa2012,yagan2012,gao2012,lee2012,brummitt2012,gomez2013diffusion,dedomenico2013random,Bianconi2013Statistical,Sola2013Centrality,pagerank2013,cozzo2012,granell2013,cozzo2013}. Moreover, despite this wealth of recent attention, we note that multiplexity was already highlighted decades ago in fields such as engineering \cite{chang1996,little2002} and sociology \cite{faust,verbrugge1979,coleman1988}.

To study multiplex and/or temporal networks systematically, it is necessary to develop a precise mathematical representation for them as well as appropriate tools to go with such a representation.  In this paper, we develop a mathematical framework for multi-layer networks using tensor algebra.  Our framework can be used to study all types of multi-layer networks (including multiplex networks, temporal networks, cognitive social structures \cite{krack1987}, multivariate networks \cite{Pattison1999Logit}, interdependent networks, etc).  To simplify exposition, we will predominantly use the language of multiplex networks in this paper, and we will thus pay particular attention to this case.

\begin{figure}[!t]
	\centering
	  \includegraphics[width=9cm]{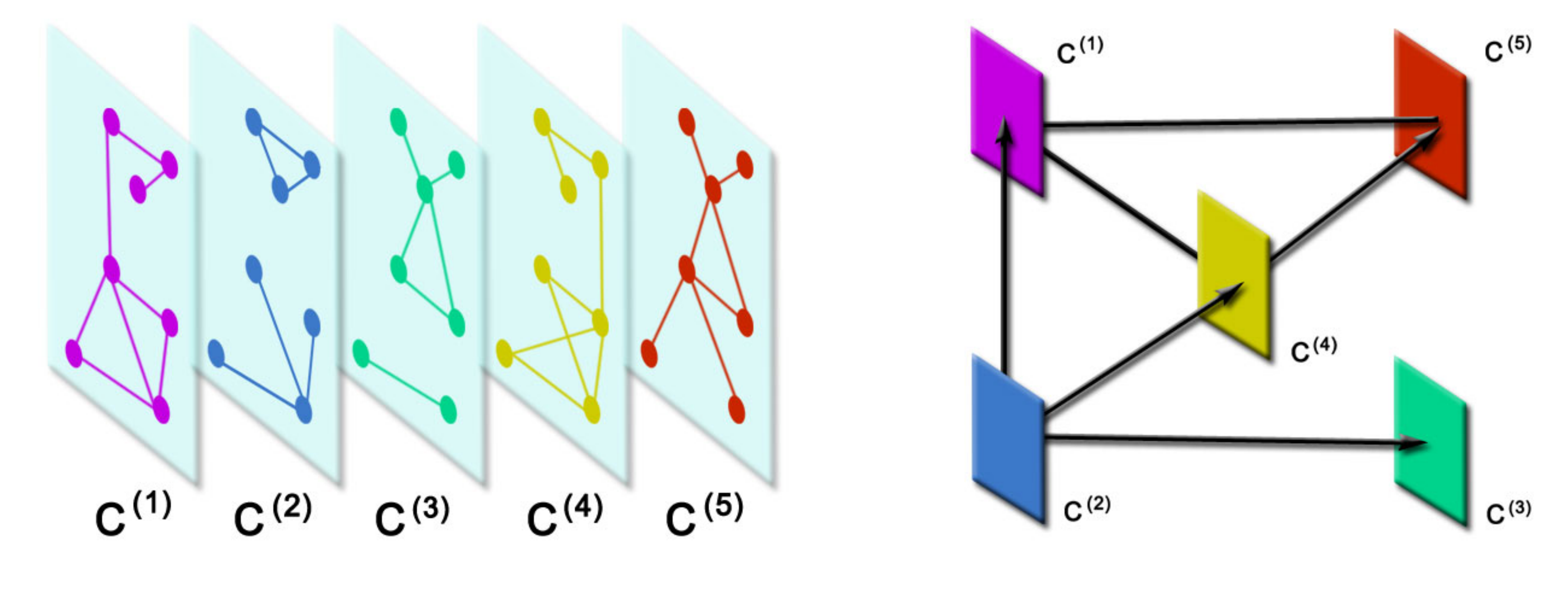}
	\caption{Schematic of a multi-layer network. When studying ordinary networks, one represents the interaction of entities (i.e., nodes) using an adjacency matrix, which encodes the intensity of each pairwise inter-relationship. However, as shown in the left panel, entities can interact in different ways (e.g., depending on the environment in which they are embedded). In this panel, each \emph{layer} of a multi-layer network corresponds to a different type of interaction (e.g., social relationships, business collaborations, etc.) and is represented by a different adjacency matrix. As shown in the right panel, the layers can also be interconnected. This introduces a new type of relationship and complicates the description of this networked system.}
        \label{fig:components}
\end{figure}

There are numerous network diagnostics for which it is desirable to develop multi-layer generalizations. In particular, we consider degree centrality, eigenvector centrality, clustering coefficients, modularity, Von Neumann entropy, and random walks. Some of these notions have been discussed previously in the context of multiplex networks \cite{mucha2010community,gomez2013diffusion,dedomenico2013random,Sola2013Centrality,granell2013,cozzo2013}.  In this paper, we define these notions for adjacency-tensor representations of multi-layer networks.  Our generalizations of these quantities are natural, and this makes it possible to compare these multi-layer quantities with their single-layer counterparts in a systematic manner. This is particularly important for the examination of new phenomena, such as multiplexity-induced correlations \cite{lee2012} and new dynamical feedbacks \cite{cozzo2012}, which arise when generalizing the usual single-layer networks. In Fig.~\ref{fig:components}, we show a schematic of multi-layer networks. In the left panel, we highlight the different interactions (edges) between entities (nodes) in different layers; in the right panel, we highlight the connections between different layers.

The remainder of this paper is organized as follows. In Section \ref{sec:monoplex}, we represent single-layer (i.e., ``monoplex") networks using a tensorial framework. We extend this to multi-layer networks in Section \ref{sec:multilayer}, and we discuss several descriptors and diagnostics for both single-layer and multi-layer networks in Section \ref{desc}. We conclude in Section \ref{conc}.

%%%%%%%

\section{Single-Layer (``Monoplex") Networks}\label{sec:monoplex}

Given a set of $N$ objects $n_{i}$ (where $i=1,2,\ldots,N$ and $N\in\mathbb{N}$), we associate to each object a state that is represented by a canonical vector in the vector space $\mathbb{R}^{N}$. More specifically, let $\mathbf{e}_{i}\equiv(0,\ldots,0,1,0,\ldots,0)^\dag$, where $\dag$ is the transposition operator, be the column vector that corresponds to the object $n_{i}$ (which we call a \emph{node}).  The $i^{\text{th}}$ component of $\mathbf{e}_{i}$ is 1, and all of its other components are $0$.

One can relate the objects $n_i$ with each other, and our goal is to find a simple way to indicate the presence and the intensity of such relationships. The most natural choice of vector space for describing the relationship is created using the tensor product (i.e., the Kronecker product) $\mathbb{R}^{N}\otimes\mathbb{R}^{N}=\mathbb{R}^{N\times N}$ \cite{mta}. Thus, $2^{\text{nd}}$-order (i.e., rank-2) canonical tensors are defined by $\mathbf{E}_{ij}=\mathbf{e}_{i}\otimes \mathbf{e}_{j}^{\dag}$ (where $i,j=1,2,\ldots,N$).
Consequently, if $w_{ij}$ indicates the intensity of the relationship from object $n_{i}$ to object $n_{j}$, we can write the \emph{relationship tensor} as
\begin{align}
\label{def:relation-tensor}
	\mathbf{W}=\sum_{i,j=1}^{N}w_{ij}\mathbf{E}_{ij}=\sum_{i,j=1}^{N}w_{ij}\mathbf{e}_{i}\otimes 	\mathbf{e}_{j}^{\dag}\,, \quad \mathbf{W}\in\mathbb{R}^{N}\otimes\mathbb{R}^{N}\,.
\end{align}
Importantly, the relationships that we have just described can be directed.  That is, the intensity of the relationship from object $n_{i}$ to object $n_{j}$ need not be the same as the intensity of the relationship from object $n_{j}$ to object $n_{i}$.

In the context of networks, $\mathbf{W}$ corresponds to an $N\times N$ weight matrix that represents the standard graph of a system that consists of $N$ nodes $n_i$.  This matrix is thus an example of an \emph{adjacency tensor}, which is the language that we will use in the rest of this paper.  To distinguish such simple networks from the more complicated situations (e.g., multiplex networks) that we discuss in this paper, we will use the term \emph{monoplex networks} to describe such standard networks, which are time-independent and possess only a single type of edge that connects its nodes.

Tensors provide a convenient mathematical representation for generalizing ordinary static networks, as they provide a natural way to encapsulate complicated sets of relationships that can also change in time \cite{mucha2010community,danichaos2013}.  Matrices are rank-2 tensors, so they are inherently limited in the complexity of the relationships that they can capture.  One can represent increasingly complicated types of relationships between nodes by considering tensors of higher order. An adjacency tensor can be written using a more compact notation that will be useful for the generalization to \emph{multi-layer} networks that we will discuss later. We will use the covariant notation introduced by Ricci and Levi-Civita in Ref.~\cite{ricci1900methodes}. In this notation, a row vector $\mathbf{a}\in\mathbb{R}^{N}$ is given by a covariant vector $a_{\alpha}$ (where $\alpha=1,2,\ldots,N$), and the corresponding contravariant vector $a^{\alpha}$ (i.e., its dual vector) is a column vector in Euclidean space.

To avoid confusion, we will use Latin letters $i,j,\ldots$ to indicate, for example, the $i^{\text{th}}$ vector, the $(ij)^{\text{th}}$ tensor, etc; and we will use Greek letters $\alpha,\beta,\ldots$ to indicate the components of a vector or a tensor. With this terminology, $e^{\alpha}(i)$ is the $\alpha^{\text{th}}$ component of the $i^{\text{th}}$ contravariant canonical vector $\mathbf{e}_{i}$ in $\mathbb{R}^{N}$, and $e_{\alpha}(j)$ is the $\alpha^{\text{th}}$ component of the $j^{\text{th}}$ covariant canonical vector in $\mathbb{R}^{N}$.

With these conventions, the adjacency tensor $\mathbf{W}$ can be represented as a linear combination of tensors in the canonical basis:
\begin{align}
\label{def:multilayer}
	W^{\alpha}_{\beta}=\sum_{i,j=1}^{N}w_{ij}e^{\alpha}(i)e_{\beta}(j)=\sum_{i,j=1}^{N}w_{ij}E^{\alpha}_{\beta}(ij)\,,
\end{align}
where $E^{\alpha}_{\beta}(ij)\in\mathbb{R}^{N\times N}$ indicates the tensor in the canonical basis that corresponds to the tensor product of the canonical vectors assigned to nodes $n_{i}$ and $n_{j}$ (i.e., it is $\mathbf{E}_{ij}$).

The adjacency tensor $W^{\alpha}_{\beta}$ is of mixed type: it is 1-covariant and 1-contravariant. This choice provides an elegant formulation for the subsequent definitions.

%%%%%%%%%%%%%%%%%%%%%%%%%%%%%%%%%%%%%
%%%%%%%%%%%%%%%%%%%%%%%%%%%%%%%%%%%%%
%%%%%%%%%%%%%%%%%%%%%%%%%%%%%%%%%%%%%

\section{Multi-Layer Networks}\label{sec:multilayer}

In the previous section, we described a procedure to build an adjacency tensor for a monoplex (i.e., single-layer) network. In general, however, there might be several types of relationships between pairs of nodes $n_{1},n_{2},\ldots,n_{N}$; and an adjacency tensor can be used to represent this situation. In other words, one can think of a more general system represented as a multi-layer object in which each type of relationship is encompassed in a single \emph{layer} $\tilde{k}$ (where $\tilde{k}=1,2,\ldots,L$) of a system.

We use the term \emph{intra-layer adjacency tensor} for the $2^{\text{nd}}$-order tensor $W^{\alpha}_{\beta}(\tilde{k})$ that indicates the relationships between nodes within the \emph{same} layer $\tilde{k}$. The tilde symbol allows us to distinguish indices that correspond to nodes from those that correspond to layers.

We take into account the possibility that a node $n_{i}$ from layer $\tilde{h}$ can be connected to any other node $n_{j}$ in any other layer $\tilde{k}$.  To encode information about relationships that incorporate multiple layers, we introduce the $2^{\text{nd}}$-order \emph{inter-layer adjacency tensor} $C^{\alpha}_{\beta}(\tilde{h}\tilde{k})$. Note that $C^{\alpha}_{\beta}(\tilde{k}\tilde{k})=W^{\alpha}_{\beta}(\tilde{k})$, so the inter-layer adjacency tensor that corresponds to the case in which a pair of layers represents the same layer $\tilde{k}$ is equivalent to the intra-layer adjacency tensor of such a layer.

Following an approach similar to that in Section \ref{sec:monoplex}, we introduce the vectors $e^{\tilde{\gamma}}(\tilde{k})$ (where $\tilde{\gamma}=1,2,\ldots,L$ and $\tilde{k}=1,2,\ldots,L$) of the canonical basis in the space $\mathbb{R}^{L}$, where the Greek index indicates the components of the vector and the Latin index indicates the $k^{\text{th}}$ canonical vector. The tilde symbol on the Greek indices allows us to distinguish these indices from the Greek indices that correspond to nodes. It is straightforward to construct the $2^{\text{nd}}$-order tensors $E^{\tilde{\gamma}}_{\tilde{\delta}}(\tilde{h}\tilde{k})=e^{\tilde{\gamma}}(\tilde{h})e_{\tilde{\delta}}(\tilde{k})$ that represent the canonical basis of the space $\mathbb{R}^{L\times L}$.

We can write the multi-layer adjacency tensor discussed early in this section using a tensor product between the adjacency tensors $C^{\alpha}_{\beta}(\tilde{h}\tilde{k})$ and the canonical tensors $E^{\tilde{\gamma}}_{\tilde{\delta}}(\tilde{h}\tilde{k})$.  This yields
\begin{align}
	M^{\alpha\tilde{\gamma}}_{\beta\tilde{\delta}} &= \sum_{\tilde{h},\tilde{k}=1}^{L}C^{\alpha}_{\beta}(\tilde{h}\tilde{k})E^{\tilde{\gamma}}_{\tilde{\delta}}(\tilde{h}\tilde{k})\nonumber\\
\label{def:multilayer2}
	&=\sum_{\tilde{h},\tilde{k}=1}^{L}\[\sum_{i,j=1}^{N}w_{ij}(\tilde{h}\tilde{k})E^{\alpha}_{\beta}(ij)\]E^{\tilde{\gamma}}_{\tilde{\delta}}(\tilde{h}\tilde{k})\nonumber\\
	&=\sum_{\tilde{h},\tilde{k}=1}^{L}\sum_{i,j=1}^{N}w_{ij}(\tilde{h}\tilde{k})\mathcal{E}^{\alpha\tilde{\gamma}}_{\beta\tilde{\delta}}(ij\tilde{h}\tilde{k})\,,
\end{align}
where $w_{ij}(\tilde{h}\tilde{k})$ are real numbers that indicate the intensity of the relationship (which may not be symmetric) between nodes $n_{i}$ in layer $\tilde{h}$ and node $n_{j}$ in layer $\tilde{k}$, and $\mathcal{E}^{\alpha\tilde{\gamma}}_{\beta\tilde{\delta}}(ij\tilde{h}\tilde{k})\equiv e^{\alpha}(i)e_{\beta}(j)e^{\tilde{\gamma}}(\tilde{h})e_{\tilde{\delta}}(\tilde{k})$ indicates the $4^{\text{th}}$-order (i.e., rank-4) tensors of the canonical basis in the space $\mathbb{R}^{N\times N\times L\times L}$.

The \emph{multi-layer adjacency tensor} $M^{\alpha\tilde{\gamma}}_{\beta\tilde{\delta}}$ is a very general object that can be used to represent a wealth of complicated relationships among nodes. In this paper, we focus on \emph{multiplex networks}. A multiplex network is a special type of multi-layer network in which the only possible types of inter-layer connections are ones in which a given node is connected to its counterpart nodes in the other layers.  In many studies of multiplex networks, it is assumed (at least implicitly) that inter-layer connections exist between counterpart nodes in all pairs of layers.  However, (i) this need not be the case; and (ii) this departs from traditional notions of multiplexity \cite{faust}, which focus on the existence of multiple types of connections and do not preclude entities from possessing only a subset of the available categories of connections.  We thus advocate a somewhat more general (and more traditional) definition of multiplex networks.

\begin{figure*}[!t]
	\centering
	\subfigure[{ Chain} ]{
	  \includegraphics[width=5.2cm]{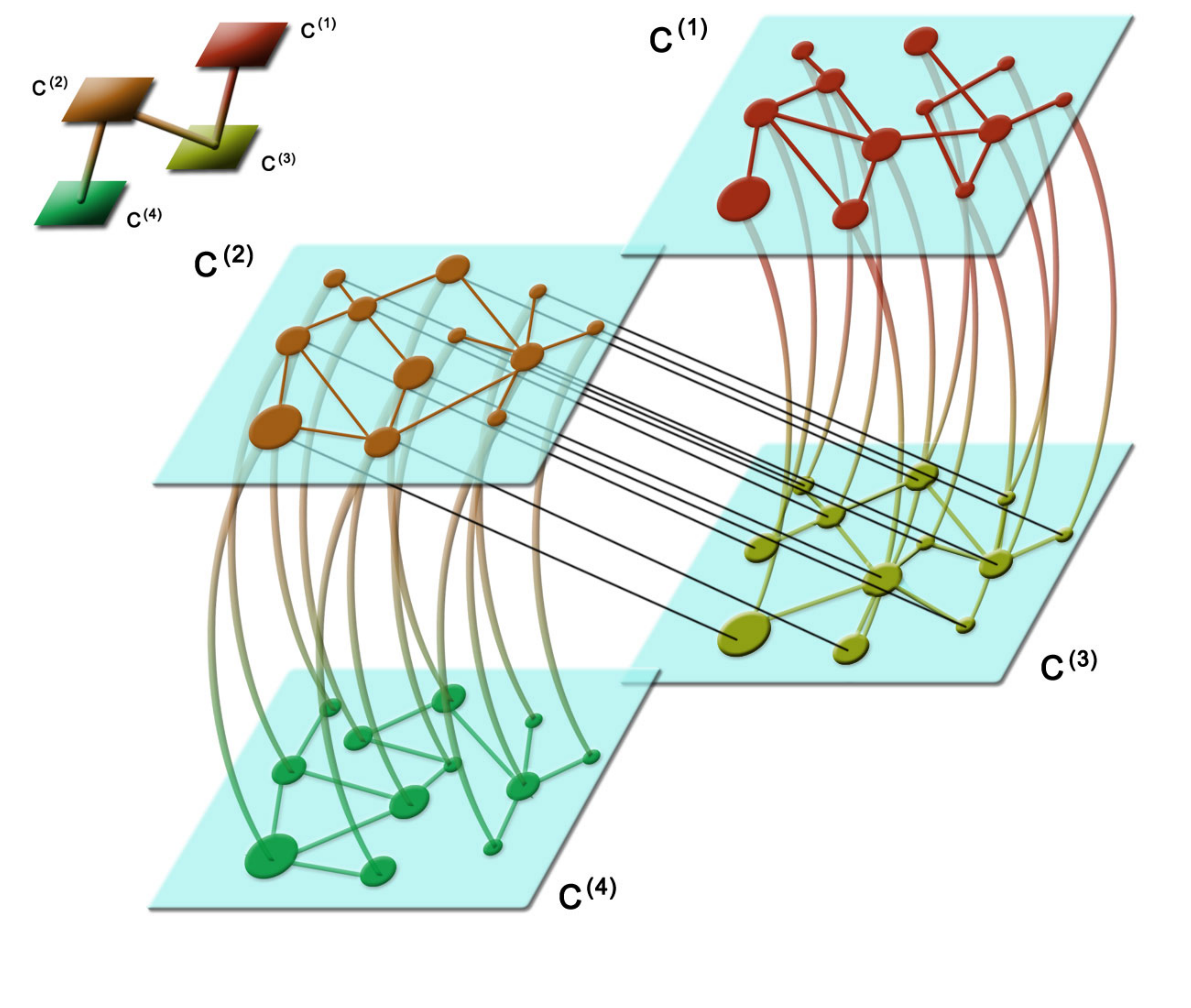}
	  }
	\subfigure[{ Star} ]{
	  \includegraphics[width=5.7cm]{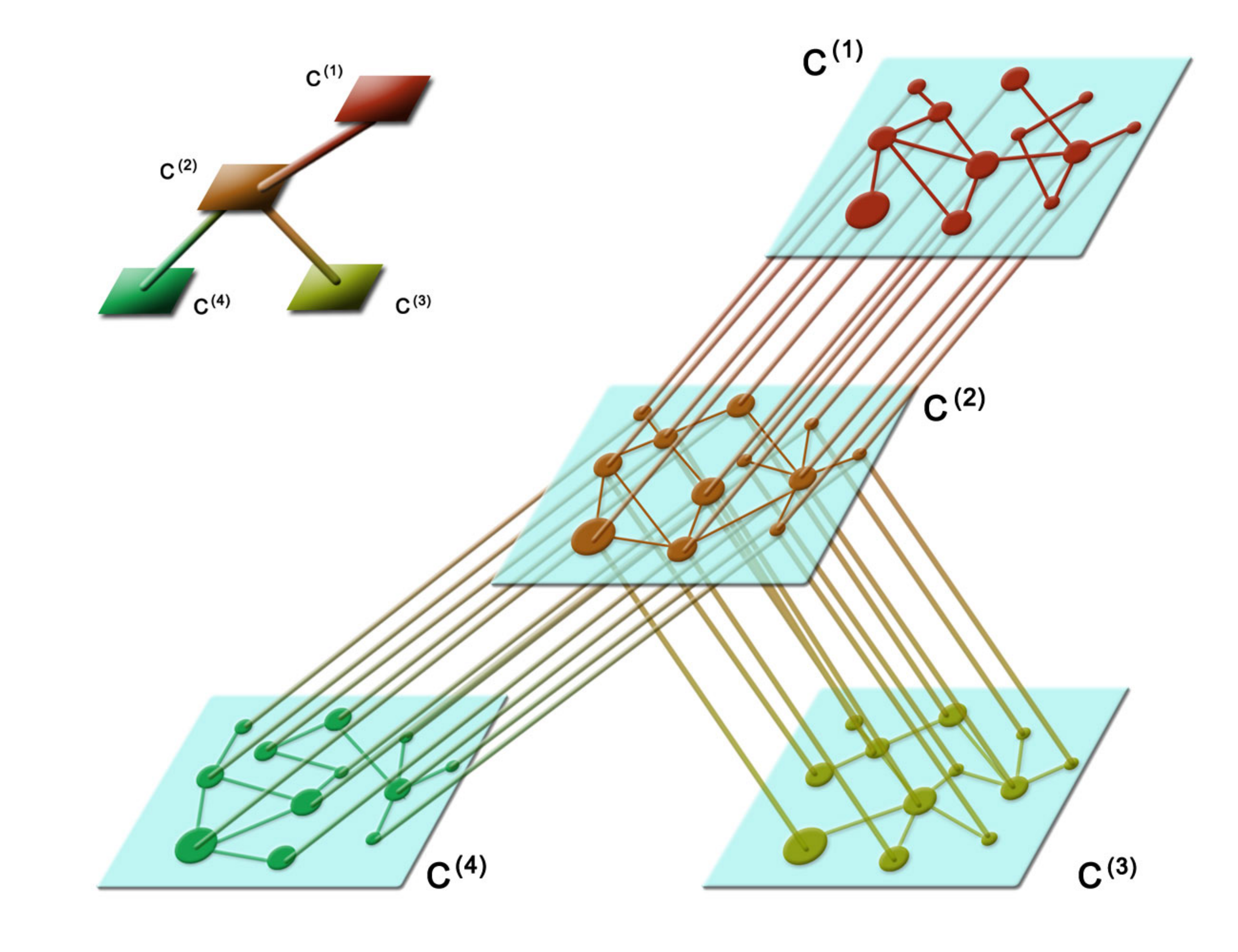}
	  }
	\subfigure[{ Lollipop} ]{
	  \includegraphics[width=5.2cm]{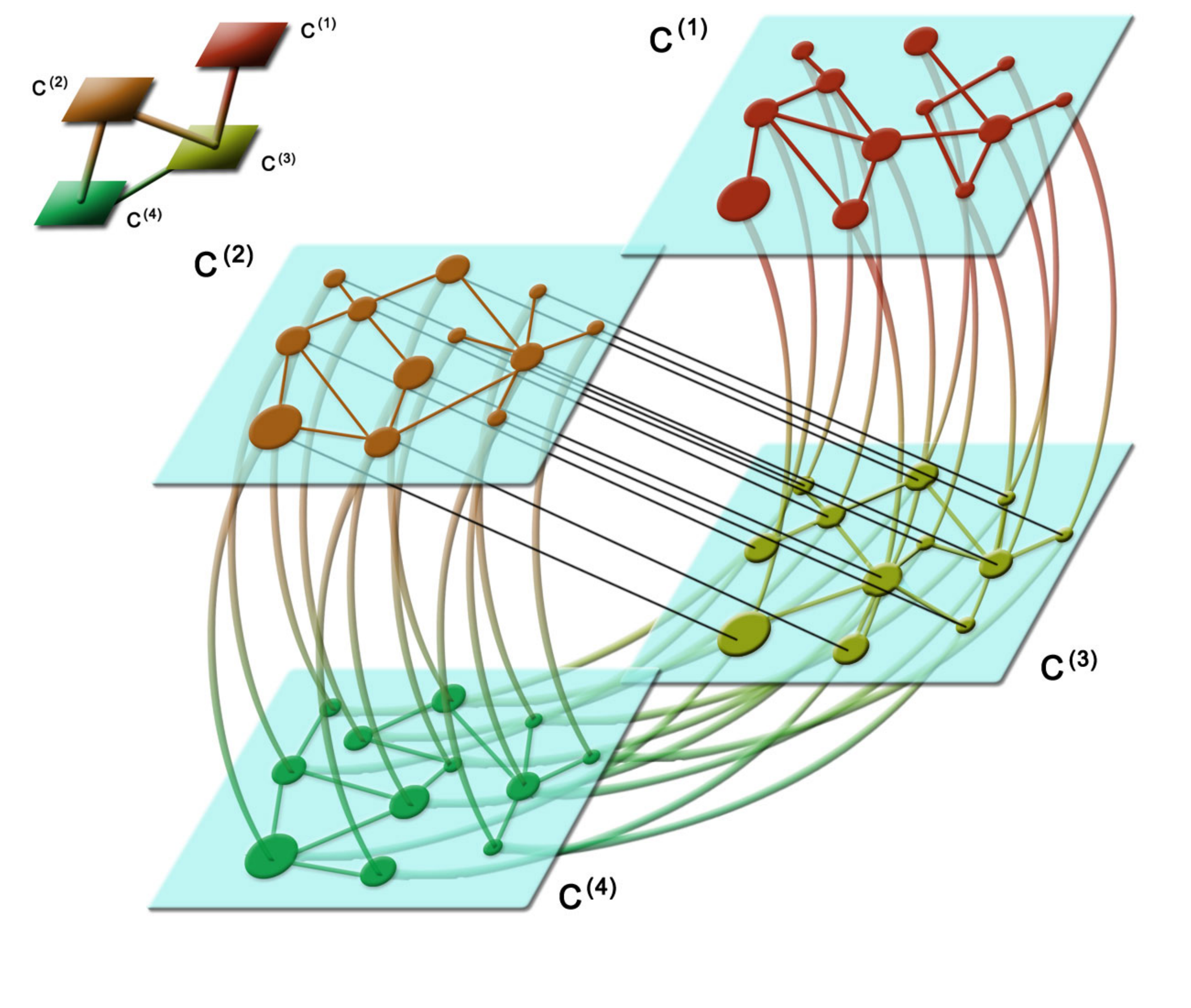}
	  }
	\caption{Schematic of multi-layer networks for three different topologies.  We show three 4-layer multiplex networks (and the corresponding network of layers as an inset in the top-left corner) and recall that each inter-layer edge connects a node with one of its counterparts in another layer.}
        \label{fig:multilayer}
\end{figure*}

When describing a multiplex network, the associated inter-layer adjacency tensor is diagonal. Importantly, connections between a node and its counterparts can have different weights for different pairs of layers, and inter-layer connections can also be different for different entities in a network \cite{mingoh2013}.  For instance, this is important for transportation networks, where one can relate the weight of inter-layer connections to the \emph{cost} of switching between a pair of transportation modes (i.e., layers).  For example, at a given station (i.e., node) in a transportation network, it takes time to walk from a train platform to a bus, and it is crucial for transportation companies to measure how long this takes \cite{horne2013}. See Fig.~\ref{fig:multilayer} for schematics of multiplex networks.

As we discussed above, entities in many systems have connections in some layers but not in others.  For example, a user of online social networks might have a Facebook account but not use Twitter.  Such an individual can thus broadcast and receive information only on a subset of the layers in the multiplex-network representation of the system.  In a transportation network, if a station does not exist in a given layer of a multi-layer network, then its associated edges also do not exist. The algebra in this paper holds for these situations without any formal modification (one simply assigns the value $0$ to associated edges), but one must think carefully about the interpretation of calculations of network diagnostics.

If there is only a single layer, there is no distinction between a monoplex network and a single-layer network, so we can use these terms interchangeably. However, the difference is crucial when studying multi-layer networks. Importantly---because it is convenient, for instance, for the implementation of computational algorithms---one can represent the multi-layer adjacency tensor $M^{\alpha\tilde{\gamma}}_{\beta\tilde{\delta}}$ as a special rank-2 tensor that one obtains by a process called \emph{matricization} (which is also known as \emph{unfolding} and \emph{flattening}) \cite{kolda2009}. The elements of $M^{\alpha\tilde{\gamma}}_{\beta\tilde{\delta}}$, which is defined in the space $\mathbb{R}^{N\times N\times L\times L}$, can be represented as an $N^{2}\times L^{2}$ or an $NL\times NL$ matrix. Flattening a multi-layer adjacency tensor can be very helpful, though of course this depends on the application in question.  Recent studies on community detection \cite{ronhovde2009multiresolution,mucha2010community}, diffusion \cite{gomez2013diffusion}, random walks \cite{dedomenico2013random}, social contagions \cite{granell2013,cozzo2013}, and clustering coefficients \cite{cozzo2013cc} on multi-layer networks have all used matrix representations of multi-layer networks for computational (and occasionally analytical) purposes.
For many years, the computational-science community has stressed the importance of developing tools for both tensors and their associated unfoldings (see the review article \cite{kolda2009} and numerous references therein) and of examining problems from these complimentary perspectives. We hope that this paper will help foster similar achievements in network science.

Another important special case of multi-layer adjacency tensors are time-dependent (i.e., ``temporal") networks.  Multi-layer representations of temporal networks have thus far tended to include only connections between a given node in one layer and its corresponding nodes in its one or two neighboring layers.  For example, the numerical calculations in Refs.~\cite{mucha2010community} and \cite{danichaos2013} only use these ``ordinal" inter-layer couplings, which causes the off-diagonal blocks of a flattened adjacency tensor to have non-zero entries only along their diagonals, even though the theoretical formulation in those papers allows more general couplings between layers.  Indeed, this restriction does not apply in general to temporal networks, as it is important for some applications to consider more general types of inter-layer couplings (e.g., if one is considering causal relationships between different nodes or if one wants to consider inter-layer coupling over a longer time horizon).  In temporal networks that are constructed from coupled time series, the individual layers in a multi-layer adjacency tensor tend to be almost completely connected (though the intra-layer edges of course have different weights).  In other cases, such as most of the temporal networks discussed in Refs.~\cite{holme2012}, there might only be a small number of non-zero connections (which represent, e.g., a small number of phone calls in a particular time window) within a single layer.

%%%%%%%%%%%%%%%%%%%%%%%%%%%%%%%%%%%%%
%%%%%%%%%%%%%%%%%%%%%%%%%%%%%%%%%%%%%
%%%%%%%%%%%%%%%%%%%%%%%%%%%%%%%%%%%%%

\section{Network Descriptors and Dynamical Processes}\label{desc}

In this section, we examine how to generalize some of the most common network descriptors and dynamical processes for multi-layer networks. First, we use our tensorial construction to show that the properties of a multi-layer network when it is made up of only a single layer reduce to the corresponding ones for a monoplex network.  We obtain these properties using algebraic operations involving the adjacency tensor, canonical vectors, and canonical tensors.  We then generalize these results for more general multi-layer adjacency tensors.

%%%%%%%%%%%%%%%%%%%%%%%%%%%%%%%%%%%%%

\subsection{Monoplex Networks}

\vspace{0.25truecm}\hspace{0.25truecm}\textbf{Degree Centrality.} Consider an undirected and unweighted network, which can be represented using the (symmetric) adjacency tensor $W^{\alpha}_{\beta}$. Define the
%\emph{unit vector}
\emph{1-vector} $u^{\alpha} = (1, \ldots, 1)^\dag \in \mathbb{R}^{N}$ whose components are all equal to 1, and let $U^{\beta}_{\alpha}=u_{\alpha}u^{\beta}$ be the $2^{\text{nd}}$-order tensor whose elements are all equal to 1. We adopt Einstein summation notation (see Appendix\,\ref{AppEinstein}) and interpret the adjacency tensor as an operator to be applied to the 1-vector. 

We thereby calculate the \emph{degree centrality vector} (or \emph{degree vector}) $k_{\beta}=W^{\alpha}_{\beta}u_{\alpha}$ in the space $\mathbb{R}^{N}$.  It is then straightforward to calculate the degree centrality of node $n_{i}$ by projecting the degree vector onto the $i^{\text{th}}$ canonical vector: $k(i)=k_{\beta}e^{\beta}(i)$. Analogously, for an undirected and weighted network, we use the corresponding weighted adjacency tensor $W^{\alpha}_{\beta}$ to define the \emph{strength centrality vector} (or \emph{strength vector}) $s_{\beta}$, which can be used to calculate the strength (i.e., weighted degree) \cite{yook2001weighted,barrat2004architecture} of each node.
With our notation, the mean degree is $\langle k\rangle=(U^{\rho}_{\rho})^{-1}k_{\beta}u^{\beta}$, the second moment of the degree is $\langle k^{2}\rangle=(U^{\rho}_{\rho})^{-1}k_{\beta}k^{\beta}$, and the variance of the degree is $\text{var}(k)=(U^{\rho}_{\rho})^{-1}k_{\beta}k^{\beta}-(U^{\rho}_{\rho})^{-2}k_{\alpha}k_{\beta}U^{\alpha\beta}$.

Directed networks are also very important, and they illustrate why it is advantageous to introduce contravariant notation.  Importantly, in-degree centrality and out-degree centrality are represented using different tensor products.  The \emph{in-degree centrality vector} is $k_{\beta}=W^{\alpha}_{\beta}u_{\alpha}$, whereas the \emph{out-degree centrality vector} is $k^{\alpha}=W^{\alpha}_{\beta}u^{\beta}$.  We then recover the usual definitions for directed networks. For example, the in-degree centrality of node $n_i$ is $k_{\text{in}}(i)=W^{\alpha}_{\beta}u_{\alpha}e^{\beta}(i)$. In directed and weighted networks, the analogous definition yields the \emph{in-strength centrality vector} and the \emph{out-strength centrality vector}. These calculations with directed networks are simple, but they illustrate that the proposed tensor algebra makes possible to develop a deeper understanding of networks, as the tensor indices are related directly to the directionality of relationships between nodes in a network.

\vspace{0.25truecm}\hspace{0.25truecm}\textbf{Clustering Coefficients.} Clustering coefficients are useful measures of transitivity in a network \cite{newman2003properties}. For unweighted and undirected networks, the local clustering coefficient of a node $n_{i}$ is defined as the number of existing edges among the set of its neighboring nodes divided by the total number of possible connections between them~\cite{watts1998collective}.  Several different definitions for local clustering coefficients have been developed for weighted and undirected networks~\cite{jp07,barrat2004architecture,ahnert2007ensemble,opsahl2009clustering} and for directed networks~\cite{fagiolo2007clustering}. Given a local clustering coefficient, one can calculate a different global clustering coefficient by averaging over all nodes. Alternatively, one can calculate a global clustering coefficient as the total number of closed triples of nodes (where all three edges are present) divided by the number of connected triples~\cite{newman2010}.

One can obtain equivalent definitions of clustering coefficients in terms of walks on a network. In standard network theory, suppose that one has an adjacency matrix ${\bf A}$ and a positive integer $m$.  Then, each matrix element $\({\bf A}^{m}\)_{ij}$ gives the number of distinct walks of length $m$ that start from node $n_{i}$ and end at node $n_{j}$. Therefore, taking $j=i$ and $m=3$ yields the number of walks of length 3 that start and end at node $n_i$. In an unweighted network without self-loops, this yields the number of distinct 3-cycles $t(i)$ that start from node $n_i$. One then calculates the local clustering coefficient of node $n_i$ by dividing $t(i)$ by the number of 3-cycles that would exist if the neighborhood of $n_i$ were completely connected.  For example, in an undirected network, one divides $t(i)$ by $k(i)\(k(i)-1\)$, which is the number of ways to select two of the neighbors of $n_i$.  In our notation, the value of $\({\bf A}^{m}\)_{ij}$ is
\begin{align}\label{cc}
	t(i,j)=W^{\alpha}_{\xi_{1}} W^{\xi_{1}}_{\xi_{2}}W^{\xi_{2}}_{\xi_{3}}\dots W^{\xi_{m-2}}_{\xi_{m-1}} W^{\xi_{m-1}}_{\beta}e_{\alpha}(i)e^{\beta}(j)\,,
\end{align}
which reduces to $t(i)=W^{\alpha}_{\rho}W^{\rho}_{\sigma}W^{\sigma}_{\beta}e_{\alpha}(i)e^{\beta}(i)$ for $j=i$ and $m=3$.
One can then define the local clustering coefficient by dividing the number of 3-cycles by the number of 3-cycles in a network for which the neighborhood of the node $n_i$ is completely connected.  This yields
\begin{align}
\label{def:nodeclus}
	c(W^{\alpha}_{\beta},i)=\frac{W^{\alpha}_{\rho}W^{\rho}_{\sigma}W^{\sigma}_{\beta}e_{\alpha}(i)e^{\beta}(i)}{W^{\alpha}_{\rho}F^{\rho}_{\sigma}W^{\sigma}_{\beta}e_{\alpha}(i)e^{\beta}(i)}\,,
\end{align}
where
\begin{align*}
  F^{\rho}_{\sigma} = U_{\sigma}^{\rho} - \delta^{\rho}_{\sigma}\,.
\end{align*}
is the adjacency tensor corresponding to a network that includes all edges except for self-loops.

To use the above formulation to calculate a global clustering coefficient of a network, we need to calculate both the total number of 3-cycles and the total number of 3-cycles that one obtains when the second step of the walk occurs in a complete network. A compact way to express this global clustering coefficient is
\begin{align}
\label{def:globalclus}
	c(W^{\alpha}_{\beta})=\frac{W^{\alpha}_{\rho}W^{\rho}_{\sigma}W^{\sigma}_{\alpha}}{W^{\alpha}_{\rho}F^{\rho}_{\sigma}W^{\sigma}_{\alpha}}\,.
\end{align}
One can define a clustering coefficient in a weighted network without any changes to Eqs.~(\ref{def:nodeclus}) and~(\ref{def:globalclus}) by assuming that $W^{\alpha}_{\beta}$ corresponds to the weighted adjacency tensor normalized such that each element of the tensor lies in the interval $[0,1]$. If weights are not defined within this range, then Eqs.~(\ref{def:nodeclus}) and~(\ref{def:globalclus}) do need to be modified.  One might also wish to modify Eq.~(\ref{def:nodeclus}) to explore generalizations of the several existing weighted clustering coefficients for ordinary networks \cite{jp07}.

We now modify Eq.~(\ref{def:globalclus}) to consider weighted clustering coefficients more generally. Let $\mathcal{N}$ be a real number that can be used to rescale the elements of the tensor.  Define $\tilde{W}^{\alpha}_{\beta}=W^{\alpha}_{\beta}/\mathcal{N}$, where one can define the normalization $\mathcal{N}$ in various ways.  For example, it can come from the maximum (so that $\mathcal{N}=\max\limits_{\alpha,\beta}\{W^{\alpha}_{\beta}\}$).
 It is straightforward to show that $c(\tilde{W}^{\alpha}_{\beta})=c(W^{\alpha}_{\beta})/\mathcal{N}$. Therefore, we redefine the global clustering coefficient $c(W^{\alpha}_{\beta})$ from Eq.~(\ref{def:nodeclus}) using this normalization:
\begin{align}
\label{def:globalclus2}
	c(W^{\alpha}_{\beta})=\mathcal{N}^{-1}\frac{W^{\alpha}_{\rho}W^{\rho}_{\sigma}W^{\sigma}_{\alpha}}{W^{\alpha}_{\rho}F^{\rho}_{\sigma}W^{\sigma}_{\alpha}}\,.
\end{align}

The same argument applies in the case of the local clustering coefficient for weighted networks. The choice of the norm in the normalization factor $\mathcal{N}$ is an important consideration.  For example, the choice $\mathcal{N}=\max\limits_{\alpha,\beta}\{W^{\alpha}_{\beta}\}$ ensures that Eq.~(\ref{def:globalclus2}) reduces to Eq.~(\ref{def:globalclus}) for unweighted networks.

\vspace{0.25truecm}\hspace{0.25truecm}\textbf{Eigenvector and Katz Centralities.} Numerous notions of centrality exist to attempt to quantify the importance of nodes (and other components) in a network \cite{faust}.  For example, a node $n_{i}$ has a high eigenvector centrality if its neighbors also have high eigenvector centrality, and the recursive nature of this notion yields a vector of centralities that satisfies an eigenvalue problem.

Let ${\bf A}$ be the adjacency matrix for an undirected network, $\mathbf{v}$ be a solution of the equation $\mathbf{A}\mathbf{v}=\lambda_1\mathbf{v}$, and $\lambda_{1}$ be the largest (``leading") eigenvalue of $\mathbf{A}$.  Thus, $\mathbf{v}$ is the leading eigenvector of ${\bf A}$, and the components of $\mathbf{v}$ give the eigenvector centralities of the nodes. That is, the eigenvector centrality of node $n_i$ is given by $v_{i}$ \cite{bonacich1972,bonacich1972b}.

In our tensorial formulation, the \emph{eigenvector centrality vector} is a solution of the tensorial equation 
\begin{equation}\label{evectens}
	W^{\alpha}_{\beta}v_{\alpha}=\lambda_{1}v_{\beta}\,, 
\end{equation}	
and $v_{\alpha}e^{\alpha}(i)$ gives the eigenvector centrality of node $n_{i}$.

For directed networks, there are two leading eigenvectors, and one needs to take into account the difference between Eq.~(\ref{evectens}) and its contravariant counterpart. Moreover, nodes with only outgoing edges have an eigenvector centrality of $0$ if the above definition is adopted. One way to address this situation is to assign a small amount $b$ of centrality to each node before calculating centrality.  One incorporates this modification of eigenvector centrality by finding the leading-eigenvector solution of the eigenvalue problem $\mathbf{v}=a \mathbf{A}\mathbf{v}+b\mathbf{1}$, where $\mathbf{1}$ is a vector in which each entry is a $1$.  This type of centrality is known as Katz centrality \cite{katz1953new}. One often chooses $b = 1$, and we note that Katz centrality is well-defined if $\lambda_{1}^{-1} > a$. Using tensorial notation, we obtain
\begin{align}
\label{def:katzcent}
	v_{\beta}=\(\delta^{\alpha}_{\beta}-aW^{\alpha}_{\beta}\)^{-1}u_{\alpha}\,.
\end{align}
To calculate Katz centrality from Eq.~(\ref{def:katzcent}), we need to calculate the tensor inverse $T^{\alpha}_{\beta}$, which satisfies the equation $T^{\alpha}_{\beta}\(\delta^{\beta}_{\sigma}-aW^{\beta}_{\sigma}\)=\delta^{\alpha}_{\sigma}$.

%%%%%%%%%%%

\vspace{0.25truecm}\hspace{0.25truecm}\textbf{Modularity.} It is often useful to decompose networks into disjoint sets (``communities") of nodes such that (relative to some null model) nodes within each community are densely connected to each other but connections between communities are sparse. Modularity is a network descriptor that can be calculated for any partition of a network into disjoint sets of nodes.  Additionally, one can attempt to algorithmically determine a partition that maximizes modularity to identify dense communities in a monoplex network.  There are many ways to maximize modularity as well as many other ways to algorithmically detect communities (see the review articles \cite{porter2009,santo2010}). We will consider Newman-Girvan modularity \cite{newman2004finding}, which is the most popular version of modularity and can be written conveniently in matrix notation \cite{newman2006modularity,newman2006PRE}. Let $S^{\alpha}_{a}$ be a tensor in $\mathbb{R}^{N\times M}$, where $\alpha$ indexes nodes and $a$ indexes the communities\footnotemark\footnotetext{The reader should be careful to not confuse the latter Latin index with the indices that we have used thus far.} in an undirected network, which can be either weighted or unweighted. The value of a component of $S^{\alpha}_{a}$ is defined to be 1 when a node belongs to a particular community and 0 when it does not. We introduce the tensor $B^{\alpha}_{\beta}=W^{\alpha}_{\beta}-k^{\alpha}k_{\beta}/\mathcal{K}$, where $\mathcal{K}=W^{\alpha}_{\beta}U^{\beta}_{\alpha}$. It follows that the modularity of a network partition is given by the scalar\footnote{Recall that swapping subscripts and superscripts (and hence covariance and contravariance) in a tensor is an implicit use of a transposition operator \cite{mta}.}
\begin{align}\label{mod-notation}
	Q = \frac{1}{\mathcal{K}}S^{a}_{\alpha}B^{\alpha}_{\beta}S^{\beta}_{a} \,.
\end{align}
To consider a general null model, we write $B^{\alpha}_{\beta}=W^{\alpha}_{\beta} - P^{\alpha}_{\beta}$, where $P^{\alpha}_{\beta}$ is a tensor that encodes the random connections against which one compares a network's actual connections.  With this general null-model tensor, modularity is also appropriate for directed networks (though, of course, it is still necessary to choose an appropriate null model).

%%%%%%%

\vspace{0.25truecm}\hspace{0.25truecm}\textbf{Von Neumann Entropy.} The study of entropy in monoplex networks has been used to help characterize complexity in networked systems \cite{sole2004information,bianconi2008entropy,anand2009entropy,bianconi2009entropy,johnson2010entropic}.  As an example, let's consider the Von Neumann entropy of a monoplex network \cite{braunstein2006laplacian}.  Recall that the Von Neumann entropy extends the Shannon (information) entropy to quantum systems.
 In quantum mechanics, the density matrix $\rho$ is a positive semidefinite operator that describes the mixed state of a quantum system, and the Von Neumann entropy of $\rho$ is defined by $\mathcal{H}(\rho)=-\text{tr}\(\rho\log_{2}\rho\)$. The eigenvalues of $\rho$ must sum to 1 to have a well-defined measure of entropy.

We also need to recall the (unnormalized) combinatorial Laplacian tensor, which is a well-known object in graph theory (see, e.g., \cite{mohar1991laplacian,godsil2001algebraic} and references therein) and is defined by $L^{\alpha}_{\beta}=\Delta^{\alpha}_{\beta}-W^{\alpha}_{\beta}$, where $\Delta^{\alpha}_{\beta}=W^{\eta}_{\gamma}u_{\eta}e^{\gamma}(\beta)\delta^{\alpha}_{\beta}$ is the strength tensor (i.e., a diagonal tensor whose elements represent the strength of the nodes). The combinatorial Laplacian is positive semidefinite, and the trace of the strength tensor is $\Delta=\Delta^{\alpha}_{\alpha}$. The eigenvalues of the density tensor $\rho^{\alpha}_{\beta}=\Delta^{-1}L^{\alpha}_{\beta}$ sum to 1, as required, and they can be used to define the Von Neumann entropy of a monoplex network using the formula
\begin{align}
\label{eq:VNentropy-monoplex}
	\mathcal{H}(W^{\alpha}_{\beta})=-\rho^{\alpha}_{\beta}\log_{2}\left[\rho^{\beta}_{\alpha}\right]\,.
\end{align}
Using the eigen-decomposition of the density tensor, one can show that the Von Neumann entropy reduces to
\begin{align}
\label{eq:VNentropy-eigenv-monoplex}
	\mathcal{H}(W^{\alpha}_{\beta})=-\Lambda^{\alpha}_{\beta}\log_{2}\left[\Lambda^{\beta}_{\alpha}\right]\,,
\end{align}
where $\Lambda^{\alpha}_{\beta}$ is the diagonal tensor whose elements are the eigenvalues of $\rho^{\alpha}_{\beta}$ (see Appendix\,\ref{appEntropy}).

\vspace{0.25truecm}\hspace{0.25truecm}\textbf{Diffusion and Random Walks.} A random walk is the simplest dynamical process that can occur on a monoplex network, and random walks can be used to approximate other types of diffusion \cite{chung1997,newman2010}.  Diffusion is also relevant for many other types of dynamical processes (e.g., for some types of synchronization \cite{arenas2008synchronization}).

Let $x_{\alpha}(t)$ denote a state vector of nodes at time $t$. The diffusion equation is
\begin{align}
	\frac{dx_{\beta}(t)}{dt}=D\[W^{\alpha}_{\beta}x_{\alpha}(t)-W^{\alpha}_{\gamma}u_{\alpha}e^{\gamma}(\beta)x_{\beta}(t)\]\,,
\end{align}
where $D$ is a diffusion constant. Recall that $s_{\gamma}=W^{\alpha}_{\gamma}u_{\alpha}$ is the strength vector and that $s_{\gamma}e^{\gamma}(\beta)x_{\beta}(t)=s_{\gamma}e^{\gamma}(\beta)\delta^{\alpha}_{\beta}x_{\alpha}(t)$. This yields the following covariant diffusion law on monoplex networks:
\begin{align}
\label{eq:diff-lapl}
	\frac{dx_{\beta}(t)}{dt}=-DL^{\alpha}_{\beta}x_{\alpha}(t)\,,
\end{align}
where $L^{\alpha}_{\beta}=W^{\eta}_{\gamma}u_{\eta}e^{\gamma}(\beta)\delta^{\alpha}_{\beta}-W^{\alpha}_{\beta}$ is the combinatorial Laplacian tensor.
The solution of Eq.~(\ref{eq:diff-lapl}) is $x_{\beta}(t)=x_{\alpha}(0)e^{-DL^{\alpha}_{\beta}t}$.

Random walks on monoplex networks \cite{chung1997,newman2010,noh2004random} have attracted considerable interest because they are both important and easy to interpret. They have yielded important insights on a huge variety of applications and can be studied analytically. For example, random walks have been used to rank Web pages \cite{pagerank1998} and sports teams \cite{callaghan2007}, optimize searches \cite{viswanathan1999optimizing}, investigate the efficiency of network navigation
\cite{yang2005exploring,da2007exploring}, characterize cyclic structures in networks \cite{rozenfeld2005statistics}, and coarse-grain networks to illuminate meso-scale features such as community structure \cite{gfeller2007spectral,rosvall2007information,lambiotte2008}.

In this paper, we consider a discrete-time random walk. Let $T^{\alpha}_{\beta}$ denote the tensor of transition probabilities between pairs of nodes, and let $p_{\alpha}(t)$ denote the vector of probabilities to find a walker at each node. Hence, the covariant master equation that governs the discrete-time evolution of the probability from time $t$ to time $t+1$ is $p_{\beta}(t+1)=T^{\alpha}_{\beta}p_{\alpha}(t)$.  One can rewrite this master equation in terms of evolving probability rates as $\dot{p}_{\beta}(t)=-\overline{L}^{\alpha}_{\beta}p_{\alpha}(t)$, where $\overline{L}^{\alpha}_{\beta}=\delta^{\alpha}_{\beta}-T^{\alpha}_{\beta}$ is the normalized Laplacian tensor. The normalized Laplacian governs the evolution of the probability-rate vector for random walks.

%%%%%%%%%%%%%%%%%%%%%%%%%%%%%%%%%%%%%

\subsection{Multi-Layer Networks}

Because of its structure, a multi-layer network can incorporate a lot of information. Before generalizing the descriptors that we discussed for monoplex networks, we discuss some algebraic operations that can be employed to extract useful information from an adjacency tensor.

\vspace{0.25truecm}\hspace{0.25truecm}\textbf{Contraction.}
Tensor contractions yield interesting quantities that are invariants when the indices are repeated (see Appendix\,\ref{AppEinstein}). For instance, one can obtain the number of nodes in a network (which is an invariant quantity) by contracting the Kronecker tensor $N=\delta^{\alpha}_{\alpha}$, where we have again used Einstein summation convention. For unweighted networks, one obtains the number of edges (which is another invariant) by calculating the scalar product between the adjacency tensor $W^{\alpha}_{\beta}$ with the dual 1-tensor $U^{\beta}_{\alpha}$ (whose components are all equal to 1).

%%%%%

\vspace{0.25truecm}\hspace{0.25truecm}\textbf{Single-Layer Extraction.} In some applications, it can be useful to extract a specific layer (e.g., the $\tilde{r}^{\text{th}}$ one) from a multi-layer network.  Using tensorial algebra, this operation is equivalent to projecting the tensor $M^{\alpha\tilde{\gamma}}_{\beta\tilde{\delta}}$ to the canonical tensor $E^{\tilde{\delta}}_{\tilde{\gamma}}(\tilde{r}\tilde{r})$ that corresponds to this particular layer. The $2^{\text{nd}}$-order canonical tensors in $\mathbb{R}^{L\times L}$ form an orthonormal basis, so the product $E^{\tilde{\gamma}}_{\tilde{\delta}}(\tilde{h}\tilde{k})E^{\tilde{\delta}}_{\tilde{\gamma}}(\tilde{r}\tilde{r})$ equals $1$ for $\tilde{h}=\tilde{k}=\tilde{r}$ and it equals $0$ otherwise. Therefore, we use Eq.~(\ref{def:multilayer2}) to write
\begin{align}
	M^{\alpha\tilde{\gamma}}_{\beta\tilde{\delta}}E^{\tilde{\delta}}_{\tilde{\gamma}}(\tilde{r}\tilde{r})=C^{\alpha}_{\beta}(\tilde{r}\tilde{r})=W^{\alpha}_{\beta}(\tilde{r})\,,
\end{align}
which is the desired adjacency tensor that corresponds to layer $\tilde{r}$. Clearly, it is possible to use an analogous procedure to extract any other tensor (e.g., ones that give inter-layer relationships). In practical applications, for example, it might be useful to extract the tensors that describe inter-layer connections between pairs of layers in the multi-layer network to compare the strengths of the coupling between them.  Another important application, which we discuss later, is the calculation of multi-layer clustering coefficients.

%%%%%

\vspace{0.25truecm}\hspace{0.25truecm}\textbf{Projected and Overlay Monoplex Networks.} In some cases, one constructs a monoplex network by aggregating multiple networks.  This is useful in many situations.  For example, the first step on studying a temporal network is often to aggregate over time.  When studying social networks, one often aggregates over different types of relationships, different communication platforms, and/or different social circles. To project a multi-layer network onto a weighted single-layer network, we multiply the corresponding tensor by the 1-tensor $U^{\tilde{\delta}}_{\tilde{\gamma}}$. The \emph{projected monoplex network} $P^{\alpha}_{\beta}=M^{\alpha\tilde{\gamma}}_{\beta\tilde{\delta}}U^{\tilde{\delta}}_{\tilde{\gamma}}$ that we obtain is
\begin{align}
	P^{\alpha}_{\beta}=\sum_{\tilde{h},\tilde{k}=1}^{L}C^{\alpha}_{\beta}(\tilde{h}\tilde{k})\,.
\end{align}

Importantly, a projected monoplex network is different from the weighted monoplex network (which one might call an \emph{overlay monoplex network}) that one obtains from a multi-layer network by summing the edges over all layers for each node.  In particular, the overlay network ignores the non-negligible contribution of inter-layer connections, which are also important for quantifying the properties of a multi-layer network. One obtains the overlay network from a multi-layer adjacency tensor by contracting the indices corresponding to the layer components:
\begin{align}
\label{def:overlaynet}
	O^{\alpha}_{\beta}=M^{\alpha\tilde{\gamma}}_{\beta\tilde{\gamma}}\,.
\end{align}

\begin{figure}[!t]
	\centering
	  \includegraphics[width=8cm]{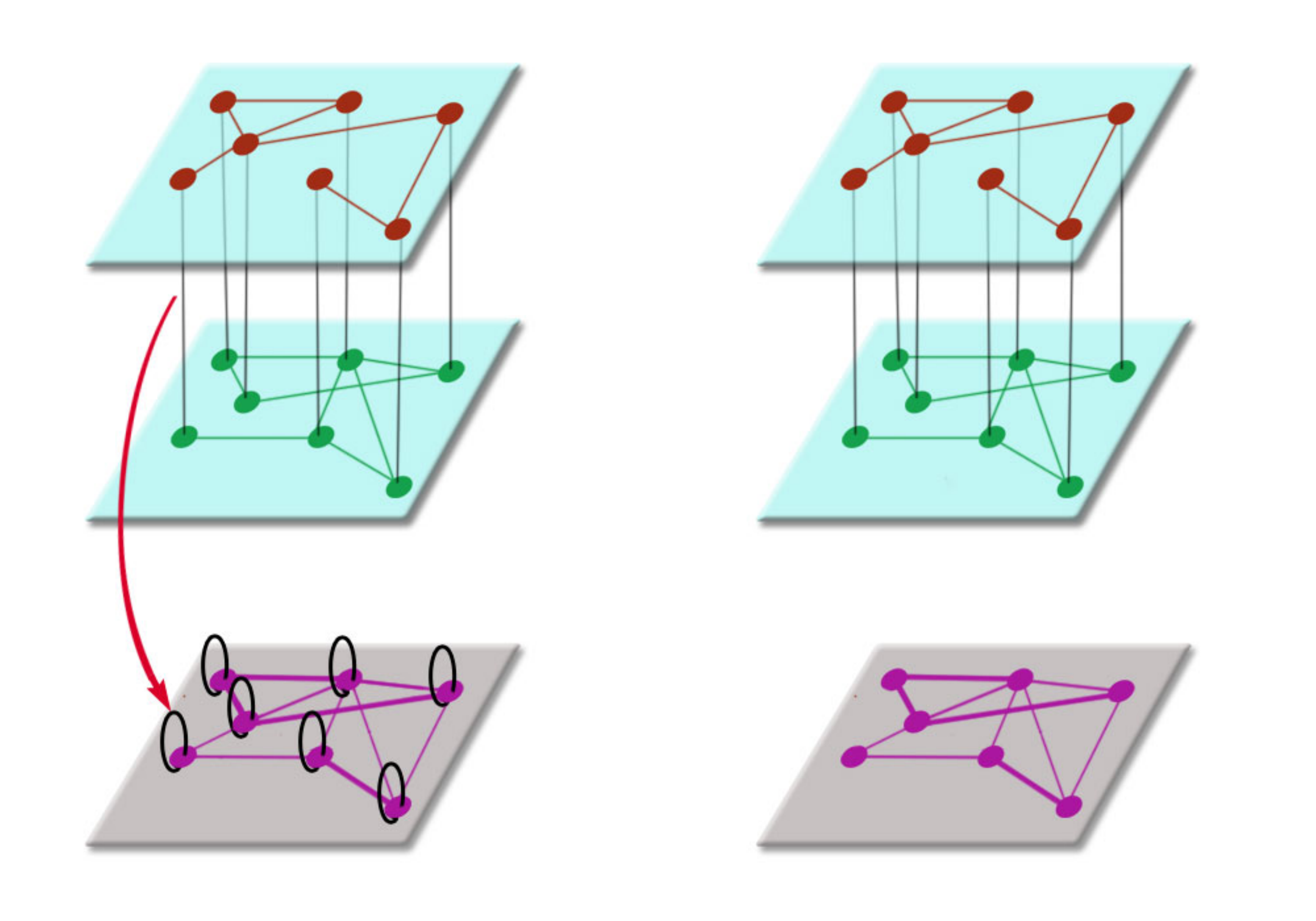}
	\caption{Schematics of (left) projected and (right) overlay monoplex networks obtained from a multi-layer network.  Both types of single-layer networks are weighted, but their edges have different weights. One obtains an overlay network using the contraction operator; this has the effect of neglecting the influence of inter-layer connections.}
        \label{fig:projoverlay}
\end{figure}

In Fig.~\ref{fig:projoverlay}, we show schematics to illustrate the difference between projected and overlay monoplex networks.

\vspace{0.25truecm}\hspace{0.25truecm}\textbf{Network of Layers.} It can be useful to construct a global observable to help understand relations between layers at a macroscopic level. For instance, suppose that two layers have more inter-connected nodes than other pairs of layers. Perhaps there are no connections across a given pair of layers. One can build a \emph{network of layers} to help understand the structure of such inter-connections.  Such a network is weighted, though the weighting procedure is application-dependent. As an example, let's consider the most intuitive weighting procedure: for each pair of layers $(\tilde{h}\tilde{k})$, we sum all of the weights in the connections between their nodes to obtain edge weights of $q_{\tilde{h}\tilde{k}}=C^{\alpha}_{\beta}(\tilde{h}\tilde{k})U_{\alpha}^{\beta}$.  For the special case of multiplex networks with unit weights between pairs of nodes in different layers, we obtain $q_{\tilde{h}\tilde{k}}=N$ if layers $\tilde{h}$ and $\tilde{k}$ are connected. The resulting weighted adjacency tensor of layers in the space $\mathbb{R}^{L\times L}$ is
\begin{align}
\label{def:layernet}
	\Psi^{\tilde{\gamma}}_{\tilde{\delta}}=\sum_{\tilde{h},\tilde{k}=1}^{L}q_{\tilde{h}\tilde{k}}E^{\tilde{\gamma}}_{\tilde{\delta}}(\tilde{h}\tilde{k})\,.
\end{align}
Hence, one can calculate $\Psi^{\tilde{\gamma}}_{\tilde{\delta}}$ from the multi-layer adjacency tensor with the formula
\begin{align}
	\Psi^{\tilde{\gamma}}_{\tilde{\delta}}=M^{\alpha\tilde{\gamma}}_{\beta\tilde{\delta}}U_{\alpha}^{\beta}\,.
\end{align}
One can then normalize the resulting tensor in a way that is appropriate for the application of interest.  For multiplex networks, for example, the most sensible normalization constant is typically the number of layers $L$.  In the insets (in the top left corners) of the panels of Fig.~\ref{fig:multilayer}, we show representations for networks of layers that correspond to three different topologies of connections between layers.

\vspace{0.25truecm}\hspace{0.25truecm}\textbf{Degree Centrality.} We now show how to compute degree centrality for multi-layer networks by performing the same projections from the case of monoplex networks using 1-tensors of an appropriate order (recall that a 1-tensor is a tensor that contains a 1 in every component). We thereby obtain a \emph{multi-degree centrality vector} $K^{\alpha}=M^{\alpha\tilde{\gamma}}_{\beta\tilde{\delta}}U^{\tilde{\delta}}_{\tilde{\gamma}}u^{\beta}$. After some algebra, we obtain
\begin{align}
	K^{\alpha}=\sum_{h,k=1}^{L}k^{\alpha}(\tilde{h}\tilde{k})\,,
\end{align}
where $k^{\alpha}(\tilde{h}\tilde{k})$ is the degree centrality vector that corresponds to connections between layers $\tilde{h}$ and $\tilde{k}$. Even in the special case of multiplex networks, it is already evident that $K^{\alpha}$ differs from the degree centrality vector that one would obtain by simply projecting all layers of a multi-layer network onto a single weighted network.

The definitions of mean degree, second moment, and variance are analogous to the corresponding monoplex network counterparts, except that one uses $K^{\alpha}$ instead of $k^{\alpha}$.

\begin{figure}[!t]
	\centering
	  \includegraphics[width=8cm]{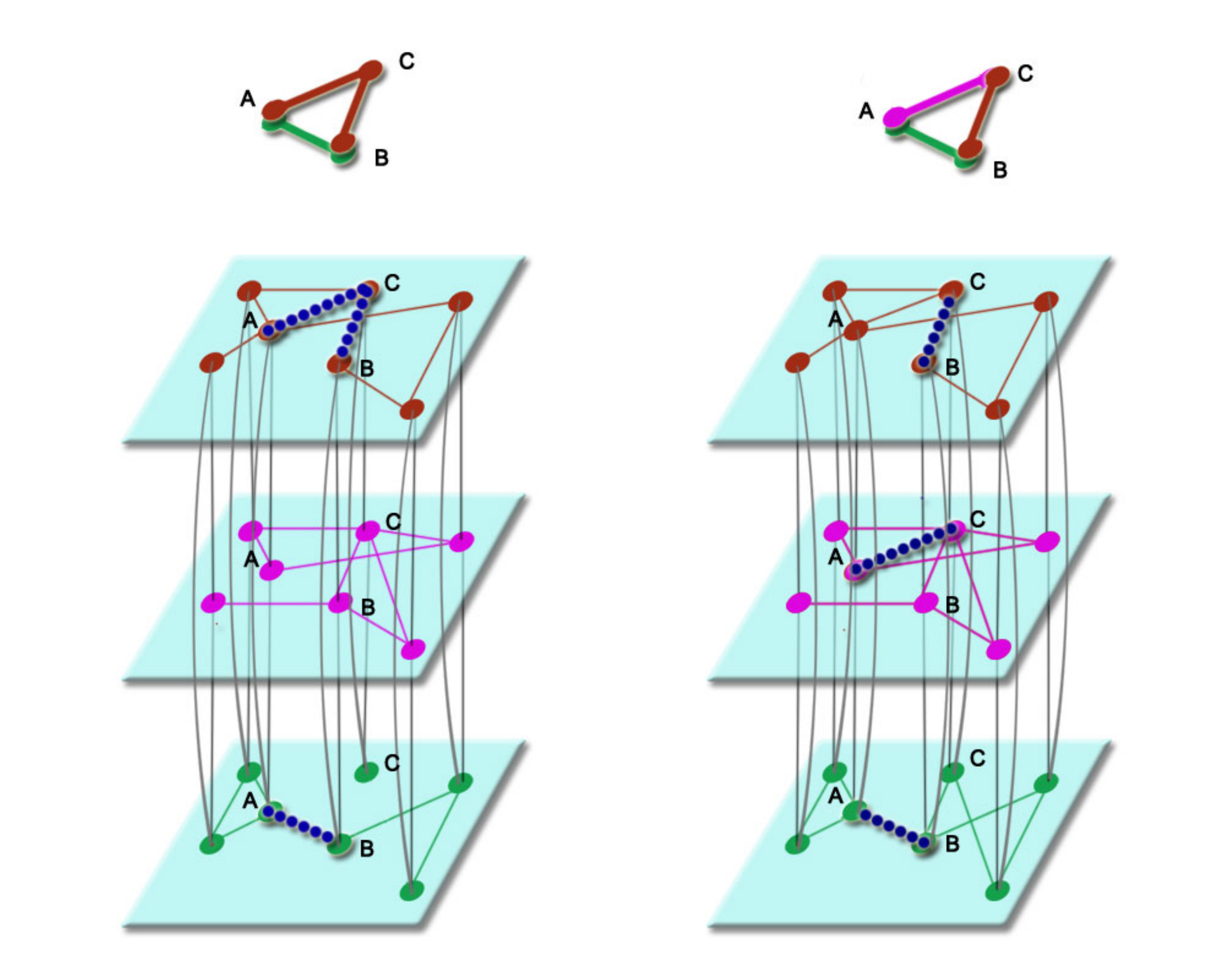}
	\caption{Schematic of closing triangles in multiplex networks.  Triangles can be closed using intra-layer connections from different layers.  In the figure, we show two different situations that can arise. For example, the left panel might represent a multi-layer social network in which nodes A and B are friends of node C but are not friends with each other (first layer), but nodes A and B still have a social tie because they work in the same company even though node $C$ does not work there (third layer). In the right panel, one might imagine that each layer corresponds to a different online social network: perhaps node B tweets about an item; node C sees node B's post on Twitter (first layer) and then posts the same item on Facebook (second layer); node A then sees this post and blogs about it (third layer), and node B reads this blog entry.}
        \label{fig:mplex-triangles}
\end{figure}

\vspace{0.25truecm}\hspace{0.25truecm}\textbf{Clustering Coefficients.}
For multi-layer networks, it is nontrivial to define a clustering coefficient using triangles as a measure of transitivity. As shown in Fig.~\ref{fig:mplex-triangles}, a closed set of three nodes might not exist on any single layer, but transitivity can still arise an a consequence of multiplexity.  In the left panel, for example, suppose that nodes A and B are friends with node C but not with each other, but that nodes A and B still have a social tie because they work at the same company (but node C does not).  Information can be passed from any one node to any other, but connections on multiple layers might be necessary for this to occur.

As with monoplex networks, we start by defining the integer power of an adjacency tensor and use a contraction operation. For instance, one can calculate the square of $M^{\alpha\tilde{\gamma}}_{\beta\tilde{\delta}}$ by constructing the $8^{\text{th}}$-order (i.e., rank-8) tensor $M^{\alpha\tilde{\gamma}}_{\beta\tilde{\delta}}M^{\rho\tilde{\epsilon}}_{\sigma\tilde{\eta}}$ and then contracting $\beta$ with $\rho$ and $\tilde{\delta}$ with $\tilde{\epsilon}$.  One computes higher powers analogously. We define a global clustering coefficient on a multi-layer adjacency tensor by generalizing Eq.~(\ref{def:globalclus}) for $4^{\text{th}}$-order tensors:
\begin{align}
\label{def:globalclusmulti}
	C(M^{\alpha\tilde{\gamma}}_{\beta\tilde{\delta}})=\mathcal{N}^{-1}\frac{ M^{\alpha\tilde{\gamma}}_{\beta\tilde{\delta}} M^{\beta\tilde{\delta}}_{\epsilon\tilde{\eta}}  M^{\epsilon\tilde{\eta}}_{\alpha\tilde{\gamma}} }{M^{\alpha\tilde{\gamma}}_{\beta\tilde{\delta}} F^{\beta\tilde{\delta}}_{\epsilon\tilde{\eta}}  M^{\epsilon\tilde{\eta}}_{\alpha\tilde{\gamma}}},
\end{align}
where we again define $F^{\beta\tilde{\delta}}_{\epsilon\tilde{\eta}}=  U^{\beta\tilde{\delta}}_{\epsilon\tilde{\eta}} -  \delta^{\beta\tilde{\delta}}_{\epsilon\tilde{\eta}}$ as the adjacency tensor of a complete multi-layer network (without self-edges). The choice of the normalization factor $\mathcal{N}$ is again arbitrary, but the choice $\mathcal{N}=\max\limits_{\alpha,\beta,\tilde{\gamma},\tilde{\delta}}\{M^{\alpha\tilde{\gamma}}_{\beta\tilde{\delta}}\}$ ensures that Eq.~(\ref{def:globalclusmulti}) is well-defined for both weighted and unweighted multi-layer networks.

The tensor contractions in Eq.~(\ref{def:globalclusmulti}) count all of the 3-cycles, including ones in which a walk goes through any combination of inter-layer and intra-layer connections. Thus, for multiplex networks with categorical layers, Eq.~(\ref{def:globalclusmulti}) counts not only fully intra-layer 3-cycles but also the inter-layer 3-cycles that are induced by the connection of nodes to their counterparts in all of the other layers.

A more traditional, and also simpler, approach to calculating a global clustering coefficient of a multi-layer network is to project it onto a single weighted network [i.e., the overlay network $O^{\alpha}_{\beta}$ defined by Eq.(\ref{def:overlaynet})] and then calculating a clustering coefficient for the resulting network. In this case, we obtain
\begin{align}\label{naive}
	c(O^{\alpha}_{\beta})=\mathcal{M}^{-1}\frac{ M^{\alpha\tilde{\gamma}}_{\beta\tilde{\gamma}} M^{\beta\tilde{\delta}}_{\epsilon\tilde{\delta}}  M^{\epsilon\tilde{\eta}}_{\alpha\tilde{\eta}} }{M^{\alpha\tilde{\gamma}}_{\beta\tilde{\gamma}} F^{\beta\tilde{\delta}}_{\epsilon\tilde{\delta}}  M^{\epsilon\tilde{\eta}}_{\alpha\tilde{\eta}}}\,,
\end{align}
where $\mathcal{M}= \max\limits_{\alpha,\beta}\{M^{\alpha\tilde{\gamma}}_{\beta\tilde{\gamma}}\}/L$. In the chosen normalization, note that we need to include a factor for the number of layers $L$, because the construction of the overlay network discards the information about the number of layers. For example, adding an empty layer to a multi-layer network does not affect the resulting overlay network, but it increases the number of possible walks that one must consider in Eq.~(\ref{naive}). In general, the clustering coefficient in Eq.~(\ref{naive}) is different from the one in  Eq.~(\ref{def:globalclusmulti}), because Eq.~(\ref{naive}) discards all inter-layer connections. However, there are some special cases when Eq.~(\ref{naive}) and Eq.~(\ref{def:globalclusmulti}) take the same value. In particular, this occurs when there is only a single layer or, more generally, when there are no inter-layer edges and all of the intra-layer networks are exactly the same.

The global clustering coefficient defined in Eq.~(\ref{naive}) sums the contributions of 3-cycles for which each step of a walk is on the same layer and those for which a walk traverses two or three layers. We decompose this clustering coefficient to separately count the contributions of 3-cycles that take place on one, two, and three layers \cite{cozzo2013cc}. To do this using our tensorial framework, we modify Eq.~(\ref{naive}) as follows:
\begin{widetext}
\begin{align}
\label{def:globalclusdecomposed}
	C(M^{\alpha\tilde{\gamma}}_{\beta\tilde{\delta}},w^\Omega)=\frac{ M^{\alpha\tilde{\gamma}}_{\beta\tilde{\delta}} M^{\beta\tilde{\rho}}_{\gamma\tilde{\sigma}}  M^{\gamma\tilde{\eta}}_{\alpha\tilde{\epsilon}}\sum\limits_{\tilde{h},\tilde{k},\tilde{l}=1}^{L}E^{\tilde{\delta}}_{\tilde{\gamma}}(\tilde{h}\tilde{h})E^{\tilde{\sigma}}_{\tilde{\rho}}(\tilde{k}\tilde{k})E^{\tilde{\epsilon}}_{\tilde{\eta}}(\tilde{l}\tilde{l})\delta_\Omega(\tilde{h},\tilde{k},\tilde{l})w^\Omega}
 {M^{\alpha\tilde{\gamma}}_{\beta\tilde{\delta}} F^{\beta\tilde{\rho}}_{\gamma\tilde{\sigma}}  M^{\gamma\tilde{\eta}}_{\alpha\tilde{\epsilon}}\sum\limits_{\tilde{h},\tilde{k},\tilde{l}=1}^{L}E^{\tilde{\delta}}_{\tilde{\gamma}}(\tilde{h}\tilde{h})E^{\tilde{\sigma}}_{\tilde{\rho}}(\tilde{k}\tilde{k})E^{\tilde{\epsilon}}_{\tilde{\eta}}(\tilde{l}\tilde{l})\delta_\Omega(\tilde{h},\tilde{k},\tilde{l})w^\Omega}\,,
\end{align}
\end{widetext}
where we have employed layer extraction operations (see our earlier discussion), $w^\Omega$ is a vector that weights the contribution of 3-cycles that span $\Omega$ layers, and $\delta_\Omega$ (where $\Omega = 1,2,3$) is a function that selects the cycles with $\Omega$ layers:
\begin{align*}\label{def:kdeltas}
	\delta_1(\tilde{h},\tilde{k},\tilde{l}) &=\delta_{\tilde{h}\tilde{l}}\delta_{\tilde{h}\tilde{k}}\delta_{\tilde{k}\tilde{l}} \,, \\
	\delta_2(\tilde{h},\tilde{k},\tilde{l}) &=(1-\delta_{\tilde{h}\tilde{l}})\delta_{\tilde{h}\tilde{k}} + (1-\delta_{\tilde{k}\tilde{l}})\delta_{\tilde{h}\tilde{k}} + (1-\delta_{\tilde{h}\tilde{k}})\delta_{\tilde{k}\tilde{l}}\,, \\
	\delta_3(\tilde{h},\tilde{k},\tilde{l}) &=(1-\delta_{\tilde{h}\tilde{l}})(1-\delta_{\tilde{k}\tilde{l}})(1-\delta_{\tilde{h}\tilde{k}})\,.
\end{align*}
We recover Eq.~(\ref{naive}) with the choice $w^\Omega=(1/3,1/3,1/3)$. (We note that $C(M^{\alpha\tilde{\gamma}}_{\beta\tilde{\delta}},w^\Omega)= C(M^{\alpha\tilde{\gamma}}_{\beta\tilde{\delta}},c w^\Omega)$ for any nonzero constant $c$, though we still normalize $w^\Omega$ by convention.)  By contrast, with the choice of $w^\Omega=(1,0,0)$, we only consider 3-cycles in which every step of a walk is on the same layer.

Importantly, we can define the weight vector $w^\Omega$ in the clustering coefficient in Eq.~(\ref{def:globalclusdecomposed}) so that it takes into account that there might be some \emph{cost} for crossing different layers. As discussed in \cite{cozzo2013cc}, one determines this ``cost'' based on the dynamics and the application of interest. For example, if one is studying a dynamical process whose time scale is very fast compared to the time scale (i.e., the cost) associated to changing layers, then it is desirable to consider the contribution from only one layer (i.e., the one in which the dynamical process occurs). For other dynamical processes, it is compulsory to also include contributions from two or three layers. To give a brief example, consider transportation at the King's Cross/St.\ Pancras station in London.  This station includes a node in a layer that describes travel via London's metro transportation system, a node in a layer for train travel within England, and a node for international train travel.  A relevant cost is then related to how long it takes to travel between different parts of the station \cite{horne2013}. One needs to consider such intra-station travel time in comparison to the schedule times of several transportation mechanisms. By contrast, there is typically very little cost associated with a person seeing information on Facebook and then posting it on Twitter.

In the above definition, the entries of $F^{\beta\tilde{\delta}}_{\epsilon\tilde{\eta}}$ are all equal to $1$ except for self-edges. Sometimes, we need to instead use a tensor $\breve{F}^{\beta\tilde{\delta}}_{\epsilon\tilde{\eta}}$ that we construct by setting some of the off-diagonal entries of $F^{\beta\tilde{\delta}}_{\epsilon\tilde{\eta}}$ to $0$.  If the original multi-layer network cannot have a particular edge, then the tensor $\breve{F}^{\beta\tilde{\delta}}_{\epsilon\tilde{\eta}}$ needs to have a $0$ in its corresponding entry. For example, this is necessary for multi-layer networks whose structural constraints forbid the existence of some nodes in certain layers or forbid certain inter- and intra-layer edges.  It is also necessary for multiplex networks, for which inter-layer edges can only exist between nodes and their counterparts in other layers. Note, however, that using the tensor $\breve{F}^{\beta\tilde{\delta}}_{\epsilon\tilde{\eta}}$ instead of $F^{\beta\tilde{\delta}}_{\epsilon\tilde{\eta}}$ influences the normalization of clustering coefficients, as it affects the set of potential 3-cycles that can exist in a multi-layer network.
%%%%%%%%%%%

\vspace{0.25truecm}\hspace{0.25truecm}\textbf{Eigenvector Centrality.} Generalizing eigenvector centrality for multi-layer network is not trivial, and there are several possible ways to do it \cite{Sola2013Centrality}.

References \cite{gomez2013diffusion,dedomenico2013random,granell2013} and \cite{cozzo2013} recently introduced the concepts of \emph{supra-adjacency} (i.e., ``super-adjacency") and \emph{supra-Laplacian} (i.e., super-Laplacian) matrices to formulate and solve eigenvalue problems in multiplex networks.  Such supra-matrices correspond to unique unfoldings of corresponding $4^{\text{th}}$-order tensors to obtain square matrices. It is worth noting that the tensorial space in which the multi-layer adjacency tensor exists is $\mathbb{R}^{N\times N\times L\times L}$, and there exists a unique unfolding---up to the $L!$ permutations of diagonal blocks of size $N\times N$ in the resulting space---that provides a square supra-adjacency tensor defined in $\mathbb{R}^{NL\times NL}$. We now exploit the same idea by arguing that a supra-eigenvector corresponds to a rank-1 unfolding of a $2^{\text{nd}}$-order ``eigentensor'' $V_{\alpha\tilde{\gamma}}$. According to this unique mapping, if $\lambda_{1}$ is the largest eigenvalue and $V_{\alpha\tilde{\gamma}}$ is the corresponding eigentensor, then it follows that
\begin{align}
	M^{\alpha\tilde{\gamma}}_{\beta\tilde{\delta}}V_{\alpha\tilde{\gamma}}=\lambda_{1} V_{\beta\tilde{\delta}}\,.
\end{align}
Therefore, similarly to monoplex networks, one can calculate the leading eigentensor $V_{\alpha\tilde{\gamma}}$ iteratively. Start with a tensor $X_{\alpha\tilde{\gamma}}(t=0)$, which we can take to be $X_{\alpha\tilde{\gamma}}(t=0) = U_{\alpha\tilde{\gamma}}$. By writing $X_{\alpha\tilde{\gamma}}(0)$ as a linear combination of the $2^{\text{nd}}$-order eigentensors and by observing that $X_{\alpha\tilde{\gamma}}(t)=\(M^{\alpha\tilde{\gamma}}_{\beta\tilde{\delta}}\)^{t}X_{\alpha\tilde{\gamma}}(0)$, one can show that $X_{\alpha\tilde{\gamma}}(t)$ is proportional to $V_{\alpha\tilde{\gamma}}$ in the $t\lto\infty$ limit. The convergence of this approach is ensured by the existence of the unfolding of $M$, since the iterative procedure is equivalent to the one applied to the corresponding supra-matrices.

We thereby obtain a multi-layer generalization of Bonacich's eigenvector centrality \cite{bonacich1972,bonacich1972b}:
\begin{align}\label{eigen-gen}
	V_{\beta\tilde{\delta}} = \lambda_{1}^{-1}M^{\alpha\tilde{\gamma}}_{\beta\tilde{\delta}}V_{\alpha\tilde{\gamma}}\,.
\end{align}

The monoplex notion of eigencentrality grants importance to a node based on its connection to other nodes.  One needs to be careful about both intra-layer and inter-layer connections when intepreting the results of calculating a multi-layer generalization of it.  For example, the intra-layer connections in one layer might be more important than those in others.  For inter-layer connections, one might ask about how much of a ``bonus" an entity earns based on its presence in multiple layers.  (This contrasts with the ``cost" that we discussed previously in the context of transportation networks.)  For instance, many web services attempt to measure the influence of people on social media by combining information from multiple online social networks, and one can choose which communication modes (i.e., layers) to include.  Moreover, by considering an overlay monoplex network or a projection monoplex network, it is possible to derive separate centrality scores for different layers.  In the above example, this would reflect the different levels of importance that somebody has on different social media.

%%%%%

\vspace{0.25truecm}\hspace{0.25truecm}\textbf{Modularity.} A multi-layer generalization of modularity was derived
in Ref.~\cite{mucha2010community} by considering random walks on networks.   Let $S^{\alpha\tilde{\rho}}_{a}$ be a tensor in $\mathbb{R}^{N\times L\times M}$, where $(\alpha,\tilde{\rho})$ indexes nodes and $a$ indexes the communities in an undirected multi-layer network, which can be either weighted or unweighted.  The value of a components of $S^{\alpha \tilde{\rho}}_{a}$ is defined to be 1 when a node belongs to a particular community and 0 when it does not. We introduce the tensor
$B^{\alpha\tilde{\rho}}_{\beta\tilde{\sigma}}=W^{\alpha\tilde{\rho}}_{\beta\tilde{\sigma}} - P^{\alpha\tilde{\rho}}_{\beta\tilde{\sigma}}$, where
$\mathcal{K}=W^{\alpha\tilde{\rho}}_{\beta\tilde{\sigma}}U^{\beta\tilde{\sigma}}_{\alpha\tilde{\rho}}$ and $P^{\alpha\tilde{\rho}}_{\beta\tilde{\sigma}}$ is a null-model tensor that encodes the random connections against which one compares a multi-layer network's actual connections. It follows that the modularity of a partition of a multi-layer network is given by the scalar
\begin{align}\label{mod-tensor}
	Q = \frac{1}{\mathcal{K}}S^{a}_{\alpha\tilde{\rho}}B^{\alpha\tilde{\rho}}_{\beta\tilde{\sigma}}S^{\beta\tilde{\sigma}}_{a}\,.
\end{align}
There are numerous choices for the null-model tensor $P^{\alpha\tilde{\rho}}_{\beta\tilde{\sigma}}$. The null models discussed in Refs.~\cite{mucha2010community,wymbs2012,danichaos2013} give special cases of the multi-layer modularity in Eq.~(\ref{mod-tensor}).

%%%%%

\vspace{0.25truecm}\hspace{0.25truecm}\textbf{Von Neumann Entropy.} To generalize the definition of Von Neumann entropy to multi-layer networks, we need to generalize the definition of the Laplacian tensor. Such an extension is not trivial because one needs to consider eigenvalues of a $4^{\text{th}}$-order tensor.

As we showed previously when generalizing eigenvector centrality, the existence of a unique unfolding into supra-matrices allows one to define and solve the eigenvalue problem
\begin{align}
	L^{\alpha\tilde{\gamma}}_{\beta\tilde{\delta}}V_{\alpha\tilde{\gamma}}=\lambda V_{\beta\tilde{\delta}}\,,
\end{align}
where $L^{\alpha\tilde{\gamma}}_{\beta\tilde{\delta}}=\Delta^{\alpha\tilde{\gamma}}_{\beta\tilde{\delta}}-M^{\alpha\tilde{\gamma}}_{\beta\tilde{\delta}}$ is the
multi-layer Laplacian tensor, $\Delta^{\alpha\tilde{\gamma}}_{\beta\tilde{\delta}}=M^{\eta\tilde{\epsilon}}_{\rho\tilde{\sigma}}U_{\eta\tilde{\epsilon}}E^{\rho\tilde{\sigma}}(\beta\tilde{\delta})\delta^{\alpha\tilde{\gamma}}_{\beta\tilde{\delta}}$ is the multi-strength tensor (i.e., the rank-$4$ counterpart of the strength tensor $\Delta^{\alpha}_{\beta}$ that we defined previously for monoplex networks), $\lambda$ is an eigenvalue, and $V_{\alpha\tilde{\gamma}}$ is its corresponding eigentensor (i.e., the unfolded rank-1 supra-eigenvector). We note that there are at most $NL$ different eigenvalues and corresponding eigentensors.

Let $\Delta=\Delta^{\alpha\tilde{\gamma}}_{\alpha\tilde{\gamma}}$ be the trace of the multi-strength tensor. The eigenvalues of the multi-layer density tensor $\rho^{\alpha\tilde{\gamma}}_{\beta\tilde{\delta}}=\Delta^{-1}L^{\alpha\tilde{\gamma}}_{\beta\tilde{\delta}}$ sum to 1, so we can use them to define the Von Neumann entropy of a multi-layer network as
\begin{align}
	\mathcal{H}(M)=-\Lambda^{\alpha\tilde{\gamma}}_{\beta\tilde{\delta}}\log_{2}\left[\Lambda^{\beta\tilde{\delta}}_{\alpha\tilde{\gamma}}\right]\,,
\end{align}
where $\Lambda^{\alpha\tilde{\gamma}}_{\beta\tilde{\beta}}$ is the diagonal tensor whose elements are the eigenvalues of $\rho^{\alpha\tilde{\gamma}}_{\beta\tilde{\delta}}$.

%%%%%%%

\vspace{0.25truecm}\hspace{0.25truecm}\textbf{Diffusion and Random Walks.} Diffusion in multiplex networks was investigated recently in Ref.~\cite{gomez2013diffusion}. A diffusion equation for multi-layer networks needs to include terms that account for inter-layer diffusion. Let $X_{\alpha\tilde{\gamma}}(t)$ denote the state tensor of nodes in each layer at time $t$. The simplest diffusion equation for a multi-layer network is then
\begin{align}
	\frac{dX_{\beta\tilde{\delta}}(t)}{dt}=M^{\alpha\tilde{\gamma}}_{\beta\tilde{\delta}}X_{\alpha\tilde{\gamma}}(t)-M^{\alpha\tilde{\gamma}}_{\rho\tilde{\sigma}}U_{\alpha\tilde{\gamma}}E^{\rho\tilde{\sigma}}(\beta\tilde{\delta})X_{\beta\tilde{\delta}}(t)\,.
\end{align}
As in the case of monoplex networks, we introduce the multi-layer combinatorial Laplacian
\begin{equation}
	L^{\alpha\tilde{\gamma}}_{\beta\tilde{\delta}}=M^{\eta\tilde{\epsilon}}_{\rho\tilde{\sigma}}U_{\eta\tilde{\epsilon}}E^{\rho\tilde{\sigma}}(\beta\tilde{\delta})\delta^{\alpha\tilde{\gamma}}_{\beta\tilde{\delta}}-M^{\alpha\tilde{\gamma}}_{\beta\tilde{\delta}}
\end{equation}
to obtain the following covariant diffusion equation for multi-layer networks:
\begin{align}
\label{eq:diff-multilapl}
	\frac{dX_{\beta\tilde{\delta}}(t)}{dt}=-L^{\alpha\tilde{\gamma}}_{\beta\tilde{\delta}}X_{\alpha\tilde{\gamma}}(t)\,.
\end{align}
The solution of Eq.~(\ref{eq:diff-multilapl}) is $X_{\beta\tilde{\delta}}(t)=X_{\alpha\tilde{\gamma}}(0)e^{-L^{\alpha\tilde{\gamma}}_{\beta\tilde{\delta}}t}$, and this provides a natural generalization of the result for monoplex networks.

\begin{figure}[!t]
	\centering
	  \includegraphics[width=8cm]{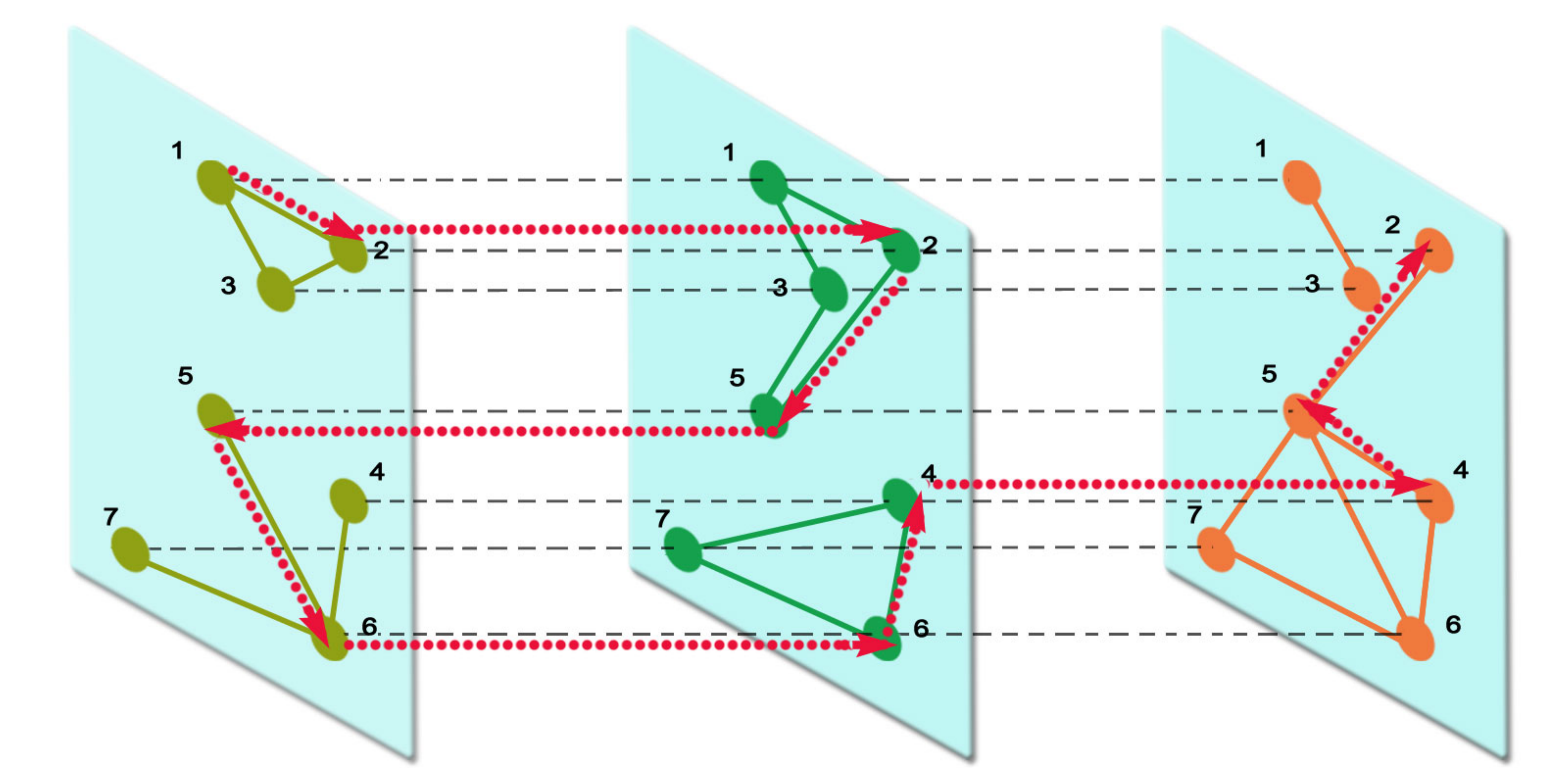}
	\caption{Schematic of a random walk (dotted trajectories) in a multiplex network. A walker can jump between nodes within the same layer, or it might switch to another layer. This illustration evinces how multiplexity allows a random walker to move between nodes that belong to different (disconnected) components on a given layer.}
        \label{fig:mplex-rw}
\end{figure}

The study of random walks is important for many applications in multi-layer networks. For instance, they were used to derive multi-layer modularity \cite{mucha2010community} and to develop optimized exploration strategies \cite{dedomenico2013random}. As we illustrate in Fig.~\ref{fig:mplex-rw}, a random walk on a multi-layer network induces nontrivial effects because the presence of inter-layer connections affects its navigation of a networked system. As with monoplex networks, we consider discrete-time random walks. Let $T^{\alpha\tilde{\gamma}}_{\beta\tilde{\delta}}$ denote the tensor of transition probabilities for jumping between pairs of nodes and switching between pairs of layers, and let $p_{\alpha\tilde{\gamma}}(t)$ be the time-dependent tensor that gives the probability to find a walker at a particular node in a particular layer. Hence, the covariant master equation that governs the discrete-time evolution of the probability from time $t$ to time $t+1$ is $p_{\beta\tilde{\delta}}(t+1)=T^{\alpha\tilde{\gamma}}_{\beta\tilde{\delta}}p_{\alpha\tilde{\gamma}}(t)$.  We rewrite this master equation in terms of evolving probability rates to obtain $\dot{p}_{\beta\tilde{\delta}}(t)=-\overline{L}^{\alpha\tilde{\gamma}}_{\beta\tilde{\delta}}p_{\alpha\tilde{\gamma}}(t)$, where $\overline{L}^{\alpha\tilde{\gamma}}_{\beta\tilde{\delta}}=\delta^{\alpha\tilde{\gamma}}_{\beta\tilde{\delta}}-T^{\alpha\tilde{\gamma}}_{\beta\tilde{\delta}}$ is the normalized Laplacian tensor.

%%%%%%%%%%%%%%%%%%%%%%%%%%%%%%%%%%%%%
%%%%%%%%%%%%%%%%%%%%%%%%%%%%%%%%%%%%%
%%%%%%%%%%%%%%%%%%%%%%%%%%%%%%%%%%%%%

\section{Conclusions and Discussion}\label{conc}

In this paper, we developed a tensorial framework to study general multi-layer networks.  We discussed the generalization of several important network descriptors---including degree centrality, clustering coefficients, eigenvector centrality, and modularity---for our multi-layer framework.  We examined different choices that one can make in developing such generalizations, and we also demonstrated how our formalism yields results for monoplex and multiplex networks as special cases.

As we have discussed in detail, our multi-layer formalism provides natural generalizations of network descriptors, and this allows systematic comparisons of multi-layer diagnostics with their single-layer counterparts.  As we have also illustrated (e.g., for global clustering coefficients), our formalism also allows systematic comparisons between different ways of generalizing familiar network concepts.
This is particularly important for the examination of new phenomena, such as multiplexity-induced correlations \cite{lee2012}, that arise when generalizing beyond the usual single-layer networks.  This occurs even for simple descriptors like degree centrality, for which the tensor indices in our formulation are related directly to the directionality of relationships between nodes in a multi-layer network.

The mathematical formalism that we have introduced can be generalized further by considering higher-order (i.e., higher-rank) tensors.  This will provide a systematic means to investigate networks that are, for example, both time-dependent and multiplex.

Our tensorial framework is an important step towards the development of a unified theoretical framework for studying networks with arbitrary complexities (including multiplexity, time-dependence, and more).  When faced with generalizing the usual adjacency matrices to incorporate a feature such as multiplexity, different scholars have employed different notation and terminology, and it is thus desirable to construct a unified framework to unifying the language for studying networks. Moreover, in addition to defining mathematical notation that simplifies the handling and generalization of previously known diagnostics on networks, a tensorial framework also offers the opportunity to unravel new properties that remain hidden when using the classical approach of adjacency matrices.
We hope to construct a proper geometrical interpretation for tensorial representations of networks and to ultimately obtain an operational theory of dynamics both on and of networks.  This perspective has led to significant advances in other areas of physics, and we believe that it will also be important for the study of networks.

%%%%%%%

\section*{Acknowledgements}

All authors were supported by the European Commission FET-Proactive project PLEXMATH (Grant No. 317614). AA, MDD, SG, and AS were also supported by the Generalitat de Catalunya 2009-SGR-838. AA also acknowledges financial support from the ICREA Academia and the James S.\ McDonnell Foundation, and SG and AA were supported by FIS2012-38266. YM was also supported by MINECO through Grants FIS2011-25167 and Comunidad de Arag\'on (Spain) through a grant to the group FENOL.  MAP acknowledges a grant (EP/J001759/1) from the EPSRC. We thank Peter Mucha and two anonymous referees for useful comments.

%%%%%

\appendix

\section{Einstein Summation Convention}\label{AppEinstein}

Einstein notation is a summation convention, which we adopt to reduce the notational complexity in our tensorial equations, that is applied to repeated indices in operations that involve tensors. For example, we use this convention in the left-hand sides of the following equations:
\begin{align}
	A^{\alpha}_{\alpha} &= \sum_{\alpha=1}^{N}A^{\alpha}_{\alpha}\,,\nonumber\\
	A^{\alpha}B_{\alpha} &= \sum_{\alpha=1}^{N}A^{\alpha}B_{\alpha}\,,\nonumber\\
	A^{\alpha}_{\beta}B^{\beta}_{\gamma}&=\sum_{\beta=1}^{N}A^{\alpha}_{\beta}B^{\beta}_{\gamma}\,,\nonumber\\
	A^{\alpha}_{\beta}B^{\beta}_{\alpha}&=\sum_{\alpha=1}^{N}\sum_{\beta=1}^{N}A^{\alpha}_{\beta}B^{\beta}_{\alpha}\,,\nonumber
\end{align}
whose right-hand sides include the summation signs explicitly.  It is straightforward to use this convention for the product of any number of tensors of any order. Repeated indices, such that one index is a subscript and the other is a superscript, is equivalent to perform a tensorial operation known as a \emph{contraction}. Contracting indices reduces the order of a tensor by 2. For instance, the contraction of the $2^{\text{nd}}$-order tensor $A^{\alpha}_{\beta}$ is the scalar $A_{\alpha}^{\alpha}$, and the $2^{\text{nd}}$-order tensors $A^{\alpha}_{\beta}B^{\beta}_{\delta}$ and $A^{\delta}_{\beta}B^{\gamma}_{\delta}$ are obtained by contracting the $4^{\text{th}}$-order tensor $A^{\alpha}_{\beta}B^{\gamma}_{\delta}$.

It is important to adopt Einstein summation on repeated indices in a way that is unambiguous. For example, $A^{\alpha}_{\alpha}=A^{\beta}_{\beta}$, but $A^{\alpha}_{\beta}B^{\beta}_{\gamma}C^{\gamma}_{\alpha}$ is \emph{not} equivalent to $A^{\alpha}_{\beta}B^{\beta}_{\beta}C^{\beta}_{\alpha}$ because of the ambiguity about the index $\beta$ in the second term of $A^{\alpha}_{\beta}B^{\beta}_{\beta}C^{\beta}_{\alpha}$.  Specifically, it is not clear if the contraction $B^{\beta}_{\beta}$ should be calculated before the product with the other tensors or vice versa.  Another situation that deserves particular attention is equations that involve a ratio between tensorial products, where one should separately apply the Einstein convention to the numerator and the denominator. Thus, one should not perform products between tensors $B^{\alpha}_{\beta}$ and $C^{\alpha}_{\beta}$ with repeated indices $\alpha$ and $\beta$ in cases like 
\begin{align}
	\frac{A^{\alpha}_{\beta}B^{\beta}_{\alpha}}{C^{\alpha}_{\beta}A^{\beta}_{\alpha}}\,.
\end{align}
This occurs, for example, in Eq.~(\ref{def:nodeclus}) in the main text.

%%%%%%%%%%%%

\section{Definition of the Tensor Exponential and Logarithm}

The exponential of a tensor $B^{\alpha}_{\beta}$ is a tensor $A^{\alpha}_{\beta}$ such that $e^{B^{\alpha}_{\beta}}=A^{\alpha}_{\beta}$. The tensor exponential is defined by the power series \cite{hirschsmale1974}
\begin{align}
	e^{B^{\alpha}_{\beta}}=\sum_{m=0}^{\infty}\frac{1}{m!}\(B^{\alpha}_{\beta}\)^{m}\,,
\end{align}
where
\begin{align}
	\(B^{\alpha}_{\beta}\)^{m} = B^{\alpha}_{\gamma_{1}}B^{\gamma_{1}}_{\gamma_{2}}B^{\gamma_{2}}_{\gamma_{3}}\dots B^{\gamma_{m-1}}_{\beta}\,.
\end{align}
A complete discussion of the properties of the tensor exponential is beyond the scope of the present paper. However, we show an example of how to calculate it for diagonalizable tensors. 

Let $B^{\alpha}_{\beta}$ be a diagonalizable tensor. In other words, there exists a diagonal tensor $D^{\alpha}_{\beta}$, whose elements are the eigenvalues of $B^{\alpha}_{\beta}$, and a tensor $J^{\alpha}_{\beta}$, whose columns are the eigenvectors of $B^{\alpha}_{\beta}$, such that $B^{\alpha}_{\beta}=J^{\alpha}_{\sigma}D^{\sigma}_{\tau}\(J^{\tau}_{\beta}\)^{-1}$. It follows that
\begin{align}
	\(B^{\alpha}_{\beta}\)^{m} &= J^{\alpha}_{\sigma}D^{\sigma}_{\tau}\(J^{\tau}_{\gamma_{1}}\)^{-1} J^{\gamma_{1}}_{\sigma_{1}}D^{\sigma_{1}}_{\tau_{1}}\(J^{\tau_{1}}_{\gamma_{2}}\)^{-1} \dots\nonumber\\
	& \qquad \dots J^{\gamma_{m-1}}_{\sigma_{m-1}}D^{\sigma_{m-1}}_{\tau_{m-1}}\(J^{\tau_{m-1}}_{\beta}\)^{-1}\nonumber\\
	&= J^{\alpha}_{\sigma}\(D^{\sigma}_{\tau}\)^{m}\(J^{\tau}_{\beta}\)^{-1}
\end{align}
and
\begin{align}
	e^{B^{\alpha}_{\beta}} &= J^{\alpha}_{\sigma}\[\sum_{m=0}^{\infty}\frac{1}{m!}\(D^{\sigma}_{\tau}\)^{m}\]\(J^{\tau}_{\beta}\)^{-1}\nonumber\\
	&=J^{\alpha}_{\sigma} e^{D^{\sigma}_{\tau}}\(J^{\tau}_{\beta}\)^{-1}\,.
\end{align}
The exponential of a diagonal tensor is the tensor obtained by exponentiating each of the diagonal elements $D^{\alpha}_{\beta}$, and it is straightforward to calculate $e^{D^{\alpha}_{\beta}}$ .

The logarithm of a tensor $A^{\alpha}_{\beta}$ is defined as the tensor $B^{\alpha}_{\beta}$ that satisfies the relation $e^{B^{\alpha}_{\beta}}=A^{\alpha}_{\beta}$. It is straightforward to show for a diagonal tensor $A^{\alpha}_{\beta}$ that 
\begin{align}
	\log \[A^{\alpha}_{\beta}\]=J^{\alpha}_{\sigma} \[\log D^{\sigma}_{\tau}\]\(J^{\tau}_{\beta}\)^{-1}\,,
\end{align}
where $D^{\alpha}_{\beta}$ is the diagonal tensor whose elements are the eigenvalues of $A^{\alpha}_{\beta}$, and $J^{\alpha}_{\beta}$ is a tensor whose columns are the eigenvectors of $A^{\alpha}_{\beta}$.

%%%%%%%%%%%%

\section{Derivation of Von Neumann Entropy}\label{appEntropy}

The Von Neumann entropy of a monoplex network is defined by Eq.~(\ref{eq:VNentropy-monoplex}). Let $\xi_{\alpha}$ be the $i^{\text{th}}$ eigenvector ($i=1,2,\ldots,N$) of the density tensor $\rho^{\alpha}_{\beta}$, and let $\Xi_{\beta}^{\alpha}$ be the tensor of eigenvectors. The density tensor is defined by rescaling the combinatorial Laplacian, so it has positive diagonal entries and non-positive off-diagonal entries.  It is positive semidefinite and has non-negative eigenvalues.  We diagonalize the density tensor to obtain $\rho^{\alpha}_{\beta}=\Xi^{\alpha}_{\sigma}\Lambda^{\sigma}_{\tau}\(\Xi^{\tau}_{\beta}\)^{-1}$, where $\Lambda^{\alpha}_{\beta}$ is a diagonal tensor whose elements are the eigenvalues of $\rho^{\alpha}_{\beta}$. These eigenvalues are equal to the eigenvalues of the combinatorial Laplacian tensor rescaled by the scalar $\Delta^{-1}$, where $\Delta=\Delta^{\alpha}_{\alpha}$ and $\Delta^{\alpha}_{\beta}$ is the strength tensor. It follows that
\begin{align}\label{c1}
	\rho^{\alpha}_{\gamma}\log_{2}\[\rho^{\gamma}_{\alpha}\] &= \(\Xi^{\alpha}_{\sigma}\Lambda^{\sigma}_{\tau}\(\Xi^{\tau}_{\gamma}\)^{-1}\)\(\Xi^{\gamma}_{\sigma}\[\log_{2}\Lambda^{\sigma}_{\tau}\]\(\Xi^{\tau}_{\alpha}\)^{-1}\)
\nonumber\\
	&= \Xi^{\alpha}_{\sigma}\Lambda^{\sigma}_{\tau}\[\log_{2}\Lambda^{\tau}_{\epsilon}\]\(\Xi^{\epsilon}_{\alpha}\)^{-1}\nonumber\\
	&=\Lambda^{\epsilon}_{\tau}\[\log_{2}\Lambda^{\tau}_{\epsilon}\]\,,
\end{align}
where we have exploited the relation $\Xi^{\alpha}_{\sigma}\(\Xi^{\epsilon}_{\alpha}\)^{-1}=\delta^{\epsilon}_{\sigma}$. We obtain Eq.\,(\ref{eq:VNentropy-eigenv-monoplex}) by multiplying both sides of Eq.~(\ref{c1}) by $-1$.

%%%%%%%%%%%%%%%%%%%%%%%%%%%%%%%%%%%%%%%%%%%
%%%%%%%%%%%%%%%%%%%%%%%%%%%%%%%%%%%%%%%%%%%
%%%%%%%%%%%%%%%%%%%%%%%%%%%%%%%%%%%%%%%%%%%
%%%%%%%%%%%%%%%%%%%%%%%%%%%%%%%%%%%%%%%%%%%

\begin{small}
\bibliography{tensor}
\end{small}

%\end{small}
%\end{multicols}
\end{document}